\documentclass[usenatbib,useAMS,onecolumn]{mn2e}

\usepackage[latin1]{inputenc}
\usepackage{psfig,epsfig}
\usepackage{amssymb}
\bibpunct{(}{)}{;}{a}{}{,}

\newcommand\etan{\eta_{\rm N}}
\newcommand\nun{\nu_{\rm N}}
\newcommand\gb[1] {   \mbox{\boldmath{$#1$}}  }
\newcommand\iffi{if and only if~}
\newcommand{\w}{kv_A}
\newcommand{\ww}{k^2v_A^2}
\newcommand{\wwww}{(kv_A)^4}
\newcommand\del{\gb{\nabla}}
\newcommand\bdot{\gb{\cdot}}
\newcommand\btimes{\gb{\times}}
\newcommand\vv{\gb{v}}
\newcommand\vu{\gb{u}}
\newcommand\B{\gb{B}}
\newcommand\Bb{\gb{b}}

\newcommand\bQ{\gb{\sigma}}
\newcommand\bJ{\gb{J}}
\newcommand\cE{\cal{E}}
\newcommand\cF{\gb{\cal{F}}}
\newcommand\bdiv[1]{\del \bdot #1}

\newcommand\hx{\gb{\hat{x}}}
\newcommand\hy{\gb{\hat{y}}}
\newcommand\hz{\gb{\hat{z}}}

\newcommand\de{\delta}
\newcommand{\dd}[2]{\frac{{\rm d} #1}{{\rm d} #2}}
\newcommand{\Dd}[2]{\frac{{\rm D} #1}{{\rm D} #2}}
\newcommand{\dpart}[2]{\frac{\partial #1}{\partial #2}}

\newcommand{\bma}[1]{ \left[ \begin{array}{#1}}
\newcommand{\ema}{\end{array} \right] } 
\newcommand{\KK}{\kappa^2}

\def\ie{i.e.~}
\def\beq{ \begin{equation} }
\def\eeq{ \end{equation} }
\def\ltsim{\mathrel{\spose{\lower.5ex\hbox{$\mathchar"218$}}
\raise.4ex\hbox{$\mathchar"13C$}}}
\def\gtsim{\mathrel{\spose{\lower.5ex\hbox{$\mathchar"218$}}
\raise.4ex\hbox{$>$}}}

\title[Exact solutions of MHD flows]
{Exact shearing box solutions of MHD flows with resistivity, viscosity
and cooling}

\author[Pierre Lesaffre \& Steven A. Balbus]
{Pierre Lesaffre$^{1}$\thanks{Email:
pierre.lesaffre@lra.ens.fr} and
Steven A. Balbus$^{1,2}$ \\
$^{1}$ Laboratoire de
Radioastronomie, 24 rue Lhomond, 75231 PARIS Cedex 05, France\\
$^{2}$ Adjunct Professor, Dept. of Astronomy, University of Virginia, Charlotesville V1 22903}

\begin{document}

\date{Received}

\maketitle

\begin{abstract}

Axisymmetric incompressible modes of the magneto-rotational
instability (MRI) with a vertical wavenumber are exact solutions of
the non-linear local equations of motion for a disk (shearing box).
They are referred to as ``channel solutions''.  Here, we generalize a
class of these solutions to include energy losses, viscous, and
resistive effects.  In the limit of zero shear, we recover the result
that torsional Alfv\'en waves are exact solutions of the non-linear
equations.  Our method allows the extension of these solutions into
the dissipative regime.

These new solutions serve as benchmarks for simulations including
dissipation and energy loss, and to calibrate numerical viscosity and
resistivity in the Zeus3D code.  We quantify the anisotropy of
numerical dissipation and compute its scaling with time and space
resolution.  We find a strong dependence of the dissipation on the
mean magnetic field that may affect the saturation state of the MRI as
computed with Zeus3D.  It is also shown that elongated grid cells
generally preclude isotropic dissipation and that a Courant time step
smaller than that which is commonly used should be taken to avoid
spurious anti-diffusion of magnetic field.

\end{abstract}
\begin{keywords}
 analytic solutions -- time-dependent -- MHD -- cooling --
resistivity -- viscosity 
\end{keywords}

\section{Introduction}

The magneto-rotational instability (MRI) \citep{BH91}, is generally
regarded as the best candidate for explaining the ``anomalous
viscosity'' that transports angular momentum through accretion disks.
The first two-dimensional numerical studies of the MRI \citep{HB92}
revealed a surprise: the flow broke along the vertical axis into two
distinct and regular sheets, which appeared as channels when
visualized in a meridional plane.  In this plane, which projects out
azimuthal motion, the perturbed velocities are nearly radial, one
flowing inward, the other outward.  This pattern was given the name of
{\it channel solution}.  It is a recurring feature of more general
three-dimensional global numerical simulations of the MRI, appearing
intermittently.  Indeed the most unstable linear modes of the MRI are
modes with a vertical wavenumber, which are, in effect, channel
solutions.

\cite{GX94} showed that these modes could be destroyed
by three-dimensional parasitic instabilities, notably the magnetic
Kelvin-Helmoltz (KH) instability.  Such studies are important for the
light they may shed on the process of the saturation of MRI
turbulence.  Resistivity and viscosity are also key to the saturation
of the MRI, as channel flows break down and fields
reconnect.  As the power of computers increases, more of these
important microphysical processes are being included in the simulations.
On the other hand, as the complexity of codes increases, fewer
benchmarks are available to check and calibrate the implementation of
numerical methods.  It is this issue that motivates this paper.

One of the key insights of \cite{GX94} was to note that the linear
modes of the MRI are {\it exact} solutions of the non-linear local
equations of incompressible magnetohydrodynamics (MHD).  This is also
seen in non-linear circularly polarised torsional Alfv\'en waves, whose
linear modes exactly satisfy the non-linear MHD equations for a
polytropic (or incompressible) gas.    In this paper, we will establish
a connection between these two types of solution.  

It is the purpose of this paper to extend the Goodman-Xu solutions to
include viscosity and resistivity.    We are also able to find solutions
that include a net heating or cooling term, provided that such
gains/losses are linear in the temperature.    We first compute the linear
MRI modes with a vertical wavenumber in the shearing box regime with viscous and resistive terms
included.    We show that the only remaining non-linear term is the total
pressure gradient (magnetic plus kinetic), and we establish a condition
for it to vanish (Section 2).  In the next two sections we investigate
less general cases without viscosity (Section 3) or without shear
(Section 4), for which we can give extended families of analytic solutions.  
In Section 5 we provide methods for finding isolated solutions under more
general assumptions.    We then present two applications of our results:
in Section 6 we benchmark a new version of the Zeus3D code with a
conservative scheme for total energy; in Section 7 we use our solutions
to compute the numerical resistivity and the numerical viscosity in the
Zeus3D code.    We discuss our results and conclude in Sections 8 and 9.



\section{General method}

\subsection{Shearing sheet}

The shearing sheet system results from a local first order expansion
of the dynamical equations of motion, with the inverse radius serving
as the small parameter.  This approach dates back to a celestial
mechanics calculation of \cite{H78}. The frame of reference rotates
at circular angular velocity $\Omega$.  Radial, azimuthal and vertical
directions are labeled by local Cartesian coordinates $x$, $y$ and
$z$.  The origin of the frame follows an unperturbed fluid element
moving in a circular orbit.  The radial logarithmic derivative of
$\Omega$ is $$A=\frac12\dd{\Omega }{\ln{R}}|_{x=0}$$
and characterises the local shear.

The fundamental dynamical equations in this rotating frame are
the mass continuity equation, 
\beq
\label{continuity}
\dpart{\rho}{t}+\bdiv{(\rho\vv)}=0,
\eeq
where $\rho$ is the mass density of the gas and $\vv$ is its velocity,
and the Navier-Stokes equation with a kinematic viscosity $\nu_V$,
\beq
\label{Euler}
\dpart{\vv}{t}+(\vv\bdot\del)\vv+2\Omega\gb{\hz \times v}+\del(2A\Omega x^2)
+\frac1{\rho}\del(p+\frac{B^2}{2})-\frac1{\rho}(\B\bdot\del)\B
=
\frac1{\rho} \bdiv(\rho \nu_V \bQ)
\eeq
where $p$ is the thermal pressure, $\B$ is the magnetic field (divided
by $2\sqrt{\pi}$) and
 $\sigma_{ij}=\frac12(\partial_i v_j+\partial_j v_i)-\frac13\partial_k v_k\delta_{ij}$
is the stress tensor. The vertical gravity is neglected.
The induction equation is
\beq
\label{induction}
\dpart{\B}{t}=\del\btimes(\vv\btimes\B-\eta_B\gb{J})
\eeq
where $\gb{J} \equiv \del\btimes\B$ and $\eta_B$ is the resistivity.  
Finally, we adopt an ideal gas equation of state with adiabatic index
$\gamma$ so that the internal energy equation reads
\beq
\label{energy}
\frac{p}{\gamma-1}\Dd{\ln (p \rho^{-\gamma})}{t} = \eta_B
J^2+\rho\nu_V \bQ\gb{:\nabla v}- \Lambda 
\eeq 
where
$\Lambda$ is the net cooling function.

Equations (\ref{continuity}) to (\ref{energy}) form the governing
system of which we seek particular solutions.  

\subsection{Incompressible vertical modes}

We seek solutions that are a single Fourier mode with vertical wavenumber $k$
and a constant vertical magnetic
field $\B_0$ superimposed on the background shear.    The solutions have the form
\beq \vv=2Ax \hy+\vu 
\quad
\B =B_0 \hz+\Bb 
\label{def1}
\eeq
with
\beq
\vu=\de \vu ~ e^{st+ikz}
\quad
\Bb=\de \Bb ~ e^{st+ikz}
\label{def2}
\eeq 

We assume $\de \vu \perp \hz$ and $\de \Bb \perp \hz$.  It is
understood that the physical solutions are obtained by taking the real
part of equations (\ref{def1}).  (In particular, note that the
$\delta$ amplitudes may be complex.  )

For this form of solution, the mass continuity equation states that the
Lagrangian derivative of $\rho$ is zero, hence  the density remains constant
along the trajectories of the fluid elements.    Thus, $\rho$ is constant
in time and space {\it provided } that it was uniform initially, which
we shall assume.  

We further assume that the pressure $p$ is initially a function of $z$ only.  
This property is also conserved in time for our particular flow.  
Upon substitution of equations (\ref{def1}) and (\ref{def2}) into equation
(\ref{Euler}), we obtain
\beq
\label{Euleru}
s\vu+2Au_x\hy+2\gb{\Omega\times u}
+\frac1{\rho}\hz\,\partial_z\left(p+\frac{(\Re[\Bb])^2}2\right)-\frac1{\rho}B_0ik\Bb
=-\nu_V k^2 \vu
\eeq
where we assume that $\nu_V$ is uniform and we use the symbols $\Re[z]$ and $\Im[z]$ to denote the real
and imaginary parts of a complex number $z$.  
Note that this equation makes sense only if the $\hz$ pressure gradient
terms vanish, a key point to which we shall return below.   

Finally, assuming a uniform resistivity $\eta_B$,
the induction equation becomes
\beq
\label{inducu}
s\Bb -2Ab_x\hy-B_0ik\vu=-\eta_B k^2 \Bb
\eeq
In what follows, we shall use the new variables 
\begin{displaymath}
\nu=k^2 \nu_V
\end{displaymath}
 and 
\begin{displaymath}
\eta=k^2 \eta_B 
\mbox{.}
\end{displaymath}

The only remaining non-linear term in the above equations is the total
pressure gradient $\partial_z p_{\rm tot}$ with $p_{\rm
tot}=p+(\Re[\Bb])^2/2$.  It is also the only term of this equation
with a $z$ component.  If, for the moment, we assume that this term is
zero, then the problem for $\vu$ and $\Bb$ is contained within
equations (\ref{Euleru}) and (\ref{inducu}) and decouples from the
energy equation.  We are left with a simple linear, incompressible
problem with viscosity and resistivity, which we now turn to solve.

The linear Euler equation may be written
\beq
\label{mau}
\mathbb{E} \vu=\frac{B_0ik}{\rho}\Bb
\eeq
with the 2$\times$2 matrix $\mathbb E$ defined by
\beq
\mathbb{E}= \bma{cc}  s+\nu   & ~-2\Omega \\
             2(\Omega+A)~ & ~s+\nu\\ \ema
\eeq
where we considered only the $x$ and $y$ components of the vectors.  
The induction equation can be rewritten similarly
\beq
\label{mab}
\mathbb{F}\Bb=B_0ik\vu
\eeq
with
\beq
\mathbb{F}= \bma{cc} s+\eta & ~0         \\
            -2A        & ~s+\eta\\ \ema
\mbox{.}
\eeq 
Operating on equation (\ref{mab}) with $\mathbb{E}$ and using 
equation (\ref{mau}) produces the linear eigenvalue problem
\beq
\label{madisp}
(\mathbb{EF}+\ww\mathbb{I})\Bb=0 ,
\eeq 
where $\mathbb{I}$ is the identity matrix and $\ww=B_0^2k^2/\rho$ where
$v_A$ is the Alfv\'en speed.  $s$ is therefore a root of the
determinant (det) polynomial $$P_{\nu,\eta}(s)={\rm
det}[\mathbb{EF}+\ww\mathbb{I}]\mbox{,}$$ which is precisely the MRI dispersion
relation.

We need compute directly only the
restricted case $P_{0,\eta}(s)$ for $\nu=0$ since the
full linear problem with parameters $(s,\nu,\eta)$ is
equivalent to the one with $(s+\nu,0,\eta-\nu)$. Hence,
$P_{\nu,\eta}(s)=P_{0,\eta-\nu}(s+\nu)$:
\beq
\label{dispeta}
P_{0,\eta}(s)=(\eta+s)^2(\kappa^2+s^2)
+2 \left[ 2A\Omega+s(s+\eta) \right] \ww+\wwww=0
\eeq
and in developed form 
\beq
P_{0,\eta}(s)=s^4 +2\eta s^3 +(\eta^2+2\ww+\KK)s^2 +
2\eta(\ww+\KK)s + \KK\eta^2+4A\Omega \ww+\wwww
\eeq
with $\KK=4\Omega(A+\Omega)$
which is identical to the form given by equation (12) in \cite{F00}.
The general dispersion relation may then be obtained by replacing
$s$ by $s+\nu$ and $\eta$ by $\eta-\nu$ 
in equation (\ref{dispeta}):
\beq
\label{disp}
P_{\nu,\eta}(s)=(\eta+s)^2\left[\kappa^2+(\nu+s)^2\right]
+2\left[2A\Omega+(s+\nu)(s+\eta)\right]\ww+\wwww=0
\eeq


\subsection{Condition for a homogeneous total pressure}

For the solutions whose form is of the previous section, the internal
energy equation becomes
\beq \label{energu} \dpart{p}{t}=(\gamma-1)
\left[\eta (\Im[\Bb])^2 +\rho \nu
(\Im[\vu])^2-\Lambda\right] \mbox{.}\eeq

Note that the $\vv\bdot\del p$ term vanishes because $\vv$
has components only along $\hx$ and $\hy$ whereas $\del p$
is along $\hz$.
The above may be rewritten as an equation for the total gas plus
magnetic pressure:
\beq
\label{dft}
\dpart{}{t}\left(p+\frac{(\Re[\Bb])^2}2\right)=\Re[s\Bb]\bdot\Re[\Bb]+
(\gamma-1)(\eta (\Im[\Bb])^2 +\rho \nu (\Im[\vu])^2-\Lambda)
\mbox{.}
\eeq

  From here on, we restrict the cooling function to be of the form
$\Lambda=-\Gamma+\alpha p$ with $\Gamma$ and $\alpha$ constant.  Since
$\rho$ is a constant, this is equivalent to taking $\Lambda$ to be a
linear function of temperature.  (In effect, this is just the leading
Taylor series expansion of $\Lambda(T)$ around an arbitrary point in temperature.)  The effect of $\Gamma$ may
be absorbed into $p_{tot}$ by adding a linear function of time with no
spatial dependence, and without loss of generality, we may set
$\Gamma=0$.  We rewrite the time evolution equation for the total
pressure accordingly:
\beq
\label{dfta}
\dpart{p_{\rm tot}}{t}+(\gamma-1)\alpha p_{\rm tot}=
\Re[s\Bb]\bdot\Re[\Bb]+
(\gamma-1)(\eta (\Im[\Bb])^2 +\rho \nu (\Im[\vu])^2
+\frac{\alpha}2\Re[\Bb]\bdot\Re[\Bb])
\mbox{.}
\eeq
The right hand side of equation (\ref{dfta})
must be independent of position
for a self-consistent solution.    
Formally, these terms may be expressed as a
spatially constant term plus 
a term of the form $\Re[a \exp(2st+2ikz)]$, with the
complex amplitude $a$ 
constant in both time and space:
\beq
\label{defa}
 a= \left[s-(\gamma-1)(\eta-\frac{\alpha}2)\right]\frac{\de b_x^2+\de b_y^2}2
-(\gamma-1)\nu\rho\frac{\de u_x^2+\de u_y^2}2
\eeq 
where we remind the reader that $\de b_x$, $\de b_y$, $\de u_x$ and $\de u_y$ are 
complex numbers.
  Our task, therefore, is
to investigate the conditions under which $a$ vanishes.  
If this can be done, the full set of equations
is reduced to the linear problem of the previous section {\rm plus}
equation (\ref{dfta}),
which now reads
\beq
\label{dfth}
\dpart{p_{\rm tot}}{t}+(\gamma-1)\alpha p_{\rm tot}=
\left[\Re[s]+(\gamma-1)(\eta+\frac{\alpha}2)\right]\frac{|\de b_x|^2+|\de b_y|^2}2
+(\gamma-1)\nu\rho \frac{|\de u_x|^2+|\de u_y|^2}2
\eeq
where $|Z|^2=(\Re[Z])^2+(\Im[Z])^2$ (modulus).
When the solution of the linear problem (\ref{Euleru})-(\ref{inducu}) 
(without the total pressure gradient) is inserted into this last
equation, one gets a simple solution for $p_{\rm tot}$ of the form :
\beq
\label{def3}
p_{\rm tot}(t)=p_1\exp(2\Re[s]t)+p_2\exp(-(\gamma-1)\alpha t)
\eeq
with
\beq
p_1=\frac1{(\gamma-1)\alpha+2\Re[s]}
\left\{
\left[\Re[s]+(\gamma-1)(\eta+\frac{\alpha}2)\right]\frac{|\de b_x|^2+|\de b_y|^2}2
+(\gamma-1)\nu\rho \frac{|\de u_x|^2+|\de u_y|^2}2
\right\}
\eeq
and
\beq
p_2=p_{\rm tot}(0) -p_1
\mbox{.}
\eeq

The above is a solution of the non-linear problem
\iffi the total pressure gradient term vanishes, hence \iffi $a=0$. 
In the case with
shear, this condition can be recast in the form of a fifth order
polynomial $Q(s)=0$. To see this, we first express the components of
$\de \Bb$ and $\de \vu$ in terms of $\de b_x$ only.    
The first row of equation (\ref{madisp}) gives $\de b_y$ as a
function of $\de b_x$:
\beq
\label{equf}
f\equiv\frac{\de b_y}{\de b_x}
=\frac{(s+\nu)(s+\eta)+4A\Omega+\ww}{2\Omega(s+\eta)}
\eeq 
Note this is valid only provided that $\Omega$ is non-zero.  If
 $\Omega=A=0$, then $\de b_x$ and $\de b_y$ are independent (see the
 case without shear in section \ref{noshear} below).

Equation (\ref{mab}) now allows us to express $\de \vu$ in terms of $\de
\Bb$ and hence $\de b_x$:

\beq
\de u_x=\frac1{ikB_0}(s+\eta)\de b_x
\eeq
and 
\beq
\de u_y=\frac1{ikB_0}\left[-2A+(s+\eta)f\right]\de b_x
\mbox{.}
\eeq
$a$ can now be rewritten
\beq
 a= \de b_x ^2 \left[s-(\gamma-1)(\eta-\frac{\alpha}2)\right]\frac12(1+f^2)
+~\de b_x ^2\,(\gamma-1)\nu\frac1{2\ww}\left[ (s+\eta)^2+(-2A+(s+\eta)f)^2\right]
\mbox{.}
\eeq
  Finally we substitute $f$ thanks to equation (\ref{equf}) and gather
quantities over the same denominator:
\beq
a=\frac{\gamma-1}{2}\,\frac{Q(s)}{(s+\eta)^2}\,{\de b_x}^2
\eeq
 where $Q(s)$ is the following polynomial in $s$:
\begin{displaymath}
Q(s)=
\left(\frac1{\gamma-1}s-\eta+\frac{\alpha}2\right)
\left[(s+\eta)^2+\frac1{4\Omega^2}\left((s+\nu)(s+\eta)+4A\Omega+\ww\right)^2\right] 
\end{displaymath}
\beq
+~\frac{\nu}{\ww}(s+\eta)^2
\left[(s+\eta)^2+\frac1{4\Omega^2}\left((s+\nu)(s+\eta)+\ww\right)^2\right]
\mbox{.}
\eeq
 The homogeneity condition $a=0$ is therefore satisfied if $s$ is a
root of $Q$. Henceforth,  we refer to $Q$ as the homogeneity
polynomial.   Note that it is generally of order 5.

  The parameters of the system are hence $\Omega$ and $A$ for the
shear, $\w$ (which actually combines $B_0$, $k$ and $\rho$) for the
magnetic field, $\nu$, $\eta$, $\gamma$ and $\alpha$ (or $\Lambda$)
for the properties of the gas.  For a given set of these parameters,
we now want to find a growth rate $s$ which satisfies the dispersion
relation (\ref{disp}) and for which the total pressure gradient
vanishes, ie: $P(s)=0$ and $Q(s)=0$.


We shall first restrict our analysis to real roots $s$ in simple cases, although
we treat the case with no shear exhaustively (see section
\ref{noshear}) .  Most of the solutions presented are
therefore standing wave solutions, except in the case without shear.
There we find circularly polarised waves and other propagating
disturbances.

\section{Inviscid solutions}

In this section we set $\nu=0$.    The homogeneity polynomial becomes

\beq
\label{Qnu0}
Q(s)=
\left(\frac1{\gamma-1}s-\eta+\frac{\alpha}2\right)
\left[(s+\eta)^2+\frac1{4\Omega^2}\left(s^2+\eta s+4A\Omega+\ww\right)^2\right] 
\mbox{.}
\eeq
We choose to find a common {\it real} root to $P$ and $Q$.  The only
 way $Q$ can have a real root is if
 $s=(\gamma-1)(\eta+\frac{\alpha}2)$ because the quantity in the
 square brackets of expression (\ref{Qnu0}) is strictly positive for $s$ real.  


\subsection{Cooling/heating}

The extra degree of freedom granted by the presence of thermal
losses/gains makes such solutions easier to find: given a real root
$s$ of $P$ (ie: a standing mode), we
need only adjust $\alpha$ to

\beq
\label{reals}
\alpha=2(\eta-\frac1{\gamma-1}s)
\mbox{.}
\eeq
The final solution is then simply given by the expressions
(\ref{def1}), (\ref{def2}) and (\ref{def3}).

\subsection{$\alpha=0$}
  From (\ref{reals}) with $\alpha=0$ we immediately see that
$s=(\gamma-1)\eta$ is required for a uniform total pressure.  
Using this in the relation $P(s)=0$, we obtain a
quadratic equation for $\eta^2$:

\beq
\gamma^2(\gamma-1)^2\eta^4
+2\gamma[(\gamma-1)\ww+2\Omega\gamma(\Omega+A)]\eta^2
+\ww(4A\Omega+\ww)=0
\eeq
Alternatively, this
 can also be viewed as a quadratic equation for $\ww$:

\beq
\label{wcond}
\wwww
+[4A\Omega+2(\gamma-1)\gamma\eta^2]\ww
+\gamma^2\eta^2[4\Omega(A+\Omega)+(\gamma-1)^2\eta^2]
=0
\mbox{.}
\eeq

Either of these equations allows a determination of the set of parameters
$(\eta,\w)$ for which there exists a solution.
  For example, in the case where $\kappa^2=4\Omega(\Omega+A)>0$, equation
(\ref{wcond}) has a real root for $\ww$ provided that
\beq 
\eta^2<\frac{\Omega A^2}{\gamma(A+\gamma\Omega)}
\eeq
which sets an upper limit on the resistivity. The request that this
root be positive sets up the additional constraint 
\beq 
\eta^2<\frac{-2A\Omega}{\gamma(\gamma-1)}
\eeq
which forces $A>0$ and sets an additional upper limit on the resistivity.
The final condition (an upper limit on resistivity) is
\beq
\eta^2<\frac{-A\Omega}{\gamma}{\rm min}
\left(\frac2{\gamma-1},\frac{-A}{A+\gamma\Omega}\right)
\mbox{.}
\eeq

\subsection{$\eta=\nu=0$, adiabatic}
In this case, the condition (\ref{wcond}) simply becomes
$\w=2\sqrt{-A\Omega}$, so $A<0$ for such a solution to
exist.  The growth rate is then $s=0$, a marginally stable mode of the
MRI.  The dispersion relation has only real roots, $s=0$ (double
root) and $s=\pm 2\sqrt{\Omega(\Omega-A)}$, but the growing and decaying modes
do not fulfill the homogeneity condition. (As mentioned
above, this solution is also valid when the constant $\Gamma$ is non
zero, which is not, strictly speaking, adiabatic.)

\section{Non rotating flow}
\label{noshear}

Here we set $A=\Omega=0$ and drop the assumption that $s$ is real.
Without rotation and shear, $\hx$ is no longer a special direction;
the direction $\hz$ is still defined by the mean field  $\B_0$.
The system is now invariant under rotation of axis $\hz$
and the eigenvectors of the linear system now depend on two independent
variables, say $\de b_x$ and $\de b_y$.
The effective dispersion relation becomes:

\beq
P(s)=R(s)^2=0 
\eeq
with

\beq
R(s)=(s+\eta)(s+\nu)+\ww
\mbox{.}
\eeq

Without shear, the homogeneity condition $a=0$ with the definition (\ref{defa})
 becomes

\beq
\label{anonrot}
\frac{a}{\gamma-1}=
\left[\frac{\alpha}2-\eta+\frac{s}{\gamma-1}+\frac{\nu}{\ww}(s+\eta)^2\right]
\frac{\de b_x^2+\de b_y^2}2
=0
\eeq
  This can be achieved if either $\de b_x^2 +\delta b_y^2=0$,
or if the factor inside
the brackets vanishes.

\subsection{Torsional Alfv\'en waves}
\label{TAW}

In this case, we simply set
$\de b_x^2+\delta b_y^2=0$ by choosing either of the circularly polarised cases $\de b_x=\pm
i\de b_y$.  Now, both roots $s$ of the dispersion relation
$R=0$ provide a possible solution:

\beq
\label{growth}
s_\pm=-\frac{\eta+\nu}2\pm\sqrt{\left(\frac{\eta-\nu}2\right)^2-\ww}
\eeq

When $\w <|{\eta-\nu}|/2$ we get two standing modes.  When $\w
>|{\eta-\nu}|/2$, $s_\pm$ have imaginary parts and the two solutions
correspond to right or left circularly polarized waves.  In
particular, when $\alpha=\nu=\eta=0$, we recover circularly polarised
torsional Alfv\'en waves which are indeed well-known solutions of
their non-linear governing equations.

\subsection{Non polarised waves without shear}
\label{nonpolar}

If $\de b_x^2 +\delta b_y^2$ is non zero, then we need to find the
common roots of $P$ and the simple quadratic

\beq
Q(s)=\frac{\alpha}2-\eta+\frac{s}{\gamma-1}+\frac{\nu}{\ww}(s+\eta)^2
\eeq
which is the factor inside the brackets of equation (\ref{anonrot}).
Since $P=R^2$, to find a common root of $P$ and $Q$ means
to find a common root of $Q$ and $R$.  For simplicity we assume
$\gamma=5/3$, but it is not much more difficult to do without this
assumption.

We detail our analysis of the common roots of $P$ and $R$ in
appendices \ref{ccomp} for complex roots and \ref{creal} for real
roots. Here, we simply summarise our result that $\alpha>\sqrt{6}|\w|$
is a necessary and sufficient condition for the existence of common
complex roots, \ie: the existence of propagating disturbances as
solutions. We are also able to give an expression for $\w$ in terms of
the other parameters of the problem in the case when there exist a
common real root, \ie: when a standing wave is solution of the
problem.

\section{Solutions with shear, resistivity, viscosity and cooling}
\label{all}

  In general, one may not be interested in  the
complete range of parameters for which a solution exists. A
benchmark calculation only needs one set of parameters.  In
that case, we may simply pick a growth rate and treat $P=0$ and $Q=0$ as
equations for $\ww$ (both quadratic).  Then
the process of finding a set of parameters that yields a solution is
greatly simplified.  As an illustration, we set $\Omega=1$, $A=-3/2$,
$\nu=1/5$ and $\eta=1/10$ and seek $\ww$ and $\alpha$ as functions of
$s$.  $P=0$ implies

\beq 
\ww=\frac1{50}\left(74-5s(3+10s) \pm5\sqrt{218-10s(11+40s)}\right)
\mbox{.}  
\eeq 
Then the equation $Q=0$ is linear in the variable $\alpha$. For
example, if we now seek a standing wave solution with the growth rate
$s=1/2$, we find

\beq
\ww=\frac{54+15\sqrt{7}}{50}\simeq1.874 
\eeq 
and from $Q(1/2)=0$
\beq
\alpha=-\frac{899569+13560\sqrt{7}}{597490}\simeq-1.566 
\mbox{.}  
\eeq
Here we have an example of an explicit benchmark with
viscosity, resistivity and heating in a shearing box.

\section{Numerical benchmarks}

  The original Zeus3D code \citep[see][]{ZI,ZII} is not written in a
fully conservative form.  In particular, equation (\ref{energy}) is
used to compute the evolution of internal energy.  In general, this
scheme leads to significant loss of total energy.  For example, if
discretisation errors lead to kinetic or magnetic energy losses, this
artificial dissipation is not reflected in viscous heating or Ohmic
heating and the total energy decreases. This energy is effectively
``radiated'' away. 

  However, when viscous and/or resistive terms are included in the
code, part or all of the total energy loss is recovered as heat and
the total energy loss is reduced. As an illustration, we ran
torsional Alfvén tests with the original Zeus3D code, and with a total
energy conserving scheme (see below).  Fig. \ref{alfven}a shows that
the total energy scheme performs much better in a case without
explicit resistive and viscous terms: the internal energy scheme
heavily distorts the wave profile \citep[this was already noted
by][]{T03}. On the other hand, fig. \ref{alfven}b shows that the
internal energy scheme with some resistivity and viscosity is
indistinguishable from the total energy scheme. Because of the finite
resolution (32 zones) there is still some numerical dissipation and
the numerical results are slightly damped compared to the analytical
solution.  The difference between the analytical solution and the
actual simulations disappears on the scale of these graphs at a
resolution of 128 zones for both these tests.

  In principle, grid based schemes cannot avoid numerical
dissipation.   However, it is possible to make numerical dissipation
look more like physical dissipation by using a conservative form for
the evolution equations.    For example, we can evolve the total energy,
and deduce the internal energy as the remainder of the mechanical
(kinetic, magnetic plus potential) energy subtracted from the total
energy.    In this case, we write the total energy equation as: 

\beq
\label{etot}
\dpart{\cE}{t}+\bdiv{\cF}=-\Lambda 
\eeq 

with 
\beq 
{\cE}=
e+\frac12\rho v^2+\frac12 b^2 +\rho \Phi 
\eeq 
where $\Phi=2A\Omega x^2$ is the tidal
potential energy, $e$ is the internal energy and 
the total energy flux is 
\beq 
\cF=
\vv\left(p+e+\frac12\rho v^2+\rho\Phi\right)
+(\B\btimes\vv)\btimes\B 
+\eta_B\bJ\btimes\B-\rho\nu_V\bQ\bdot\vv
\mbox{.}  
\eeq

  We have implemented this in the Zeus3D code.  This is similar to the
work of \cite{T03} and \cite{H06}, but we also include Ohmic stresses
$\eta_B\bJ\btimes\B$ and tidal potential energy flux $\vv\rho\Phi$.
The question arises at what stage of the calculation each term of the
total energy flux $\cF$ should be evaluated.  We ran various
benchmarks (torsional Alfvén waves and standing mode solutions
presented in the previous sections) and varied the order with which
the fluxes were computed. We noted that it is crucial to compute each
flux term simultaneously with its corresponding source or transport
term.  In particular, it is critical to compute
$(\B\btimes\vv)\btimes\B$ using the time centred values for $\B$ and
$\vv$ computed with the method of characteristics \citep[MOC,
see][]{ZII}.  On the other hand, the kinetic energy flux should not be
directionally split the way the momentum transport step is.

  Finally, it is much better to join the tidal flux to the density
transport term and not to the tidal force source term. We now
illustrate how we used our analytical solutions to prove this last
point (see fig.  \ref{mri}). We used our code with two slightly
different versions in order to reproduce the analytical solution
presented in section \ref{all}. The first version (dotted lines on
fig. \ref{mri}) would compute the tidal potential flux $\rho \Phi \vv$
at the same time as the tidal source term. The second version (dashed
lines on fig. \ref{mri}) would compute this flux jointly with the
transport step.  In the first version, the resulting total pressure
gradient is not flat and the magnetic energy loses its low $z$/high
$z$ symmetry.  The second version retains the correct symmetry and
displays a flat pressure profile.  Note however that in both
computations the average total pressure and the magnetic pressure are
slightly lower than the analytical solution.  Both simulations shown
are for cubic boxes of 32 zones aside and higher resolution improves
the magnetic pressure more efficiently than the total pressure.

  The final scheme we adopted was to compute $\cF$ in five distinct
steps:
\begin{itemize}
\item $\vv p$ is first computed using an upwinded pressure computed
at the same time as the pressure gradient source,
\item the viscous term is computed at the same time as the viscous forces,  
\item the remainder of the flux $\vv\left(e+\frac12\rho v^2+\rho\Phi\right)$  is added after the hydrodynamical
transport term,
\item the resistive term is computed along with the resistive electromotive
force,
\item the $(\B\btimes\vv)\btimes\B$ term is finally computed with the MOC advanced
$\vv$ and $\B$ which are used for the constrained transport of $\B$.
\end{itemize}
  
  Along this process, we evolve the internal energy $e$ thanks to
 equation (\ref{energy}). In particular, this provides an advanced
 estimate for $e$ in the flux term $\vv(e+\dots)$. At the end of these
 steps, we compute $\cE^*$ with the updated values of all variables.
 We then use equation (\ref{etot}) with $\Lambda=0$ to advance the
 total energy to its new value $\cE$. If the internal energy scheme
 was perfect, we would have $\cE^*=\cE$. However, this is almost
 never the case and a correction $\cE-\cE^*$ needs to be applied to
 $e$ in order to conserve total energy. We deduce the rate of
 correction of internal energy $\dot{e}=(\cE-\cE^*)/\Delta t$ where
 $\Delta t$ is the length of the time step.  The internal energy is
 finally updated with $\Lambda\neq 0$ and $\dot{e}$ thanks to an
 isochore heating/cooling step.

  

\begin{figure}
\centerline{
\begin{tabular}{cc}
\psfig{file=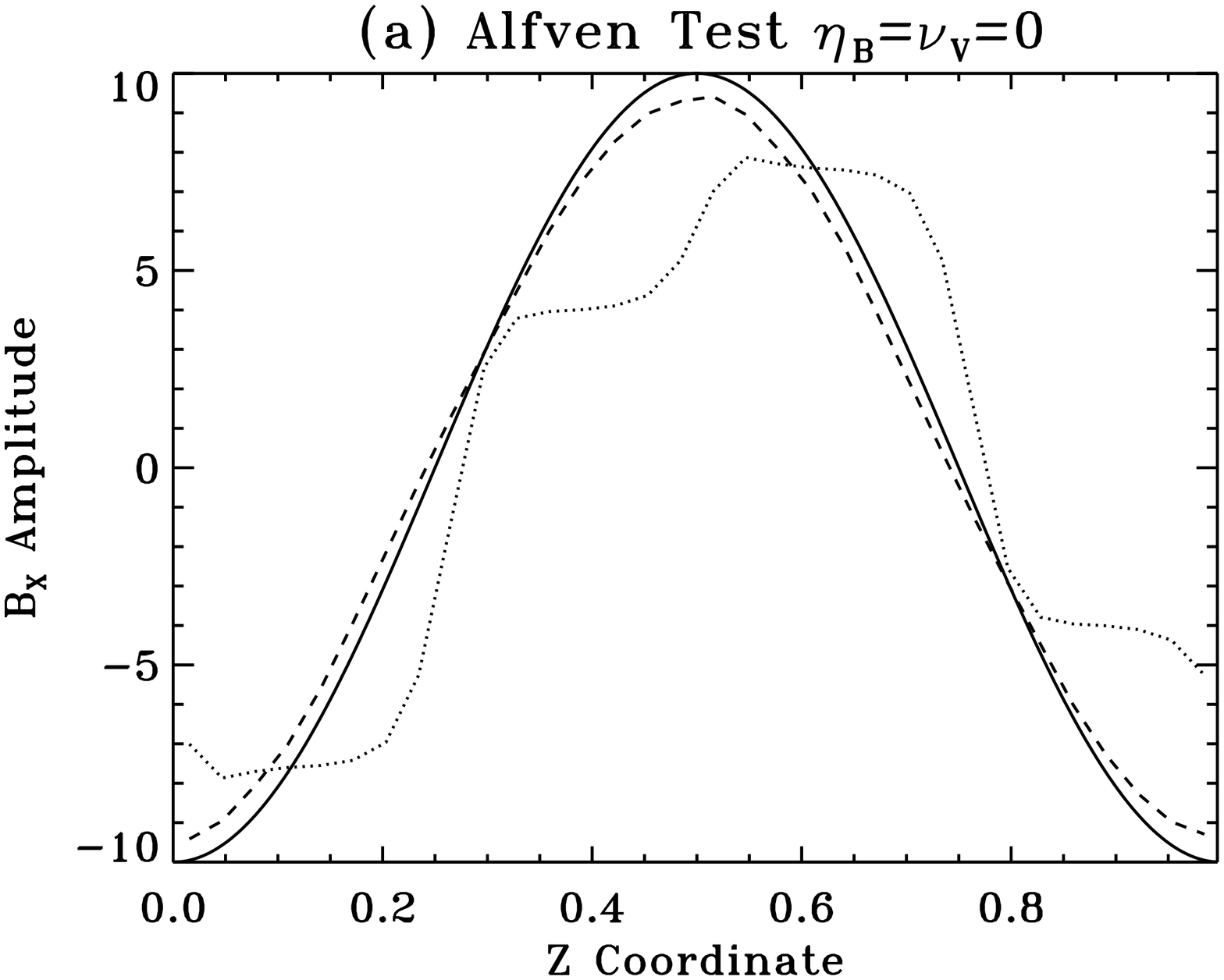,width=6cm,angle=+90}&
\psfig{file=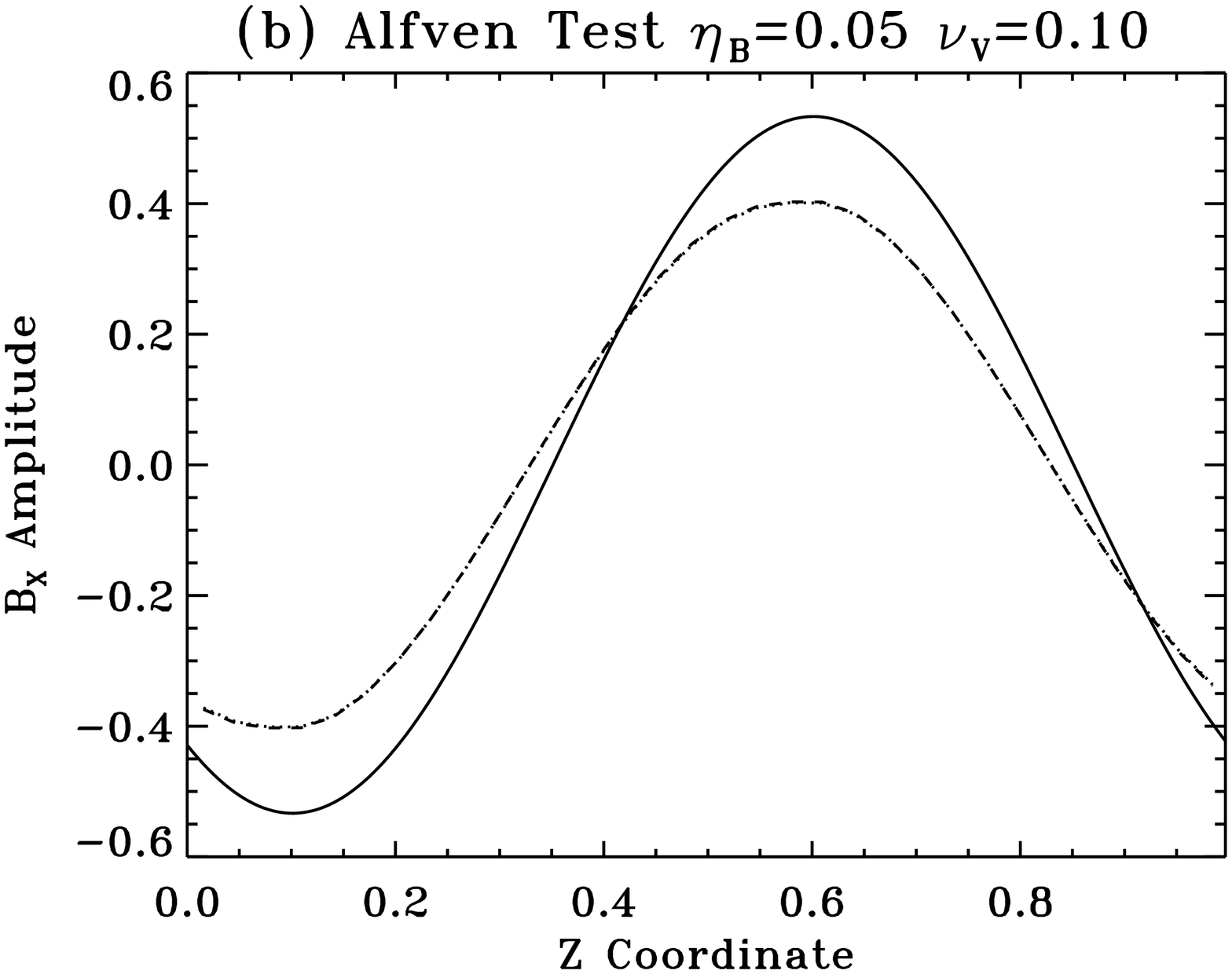,width=6cm,angle=+90}\\
\end{tabular}
} 
 \caption{Torsional Aflvén waves tests.  {\it Solid line:} analytical
solution, {\it dotted line:} internal energy scheme, {\it dashed
line:} total energy scheme.  Parameters are $\Omega=A=\alpha=0$,
$\gamma=5/3$, $B_0=\de b_x=10$, $k=2\pi$.  The simulation box is a
cube of 32 zones aside and has physical length 1.  The plots show
snapshots of the azimuthal magnetic field evaluated on a vertical
line.  {\it (a) Left panel:} $\nu_V=\eta_B=0$, after $3$ oscillations.
  {\it (b) Right panel:} $\nu_V=0.1$ and
$\eta_B=0.05$ at a time correponding to $9.9$ oscillations.   }
\label{alfven}
\end{figure}

\begin{figure}
\centerline{
\begin{tabular}{cc}
\psfig{file=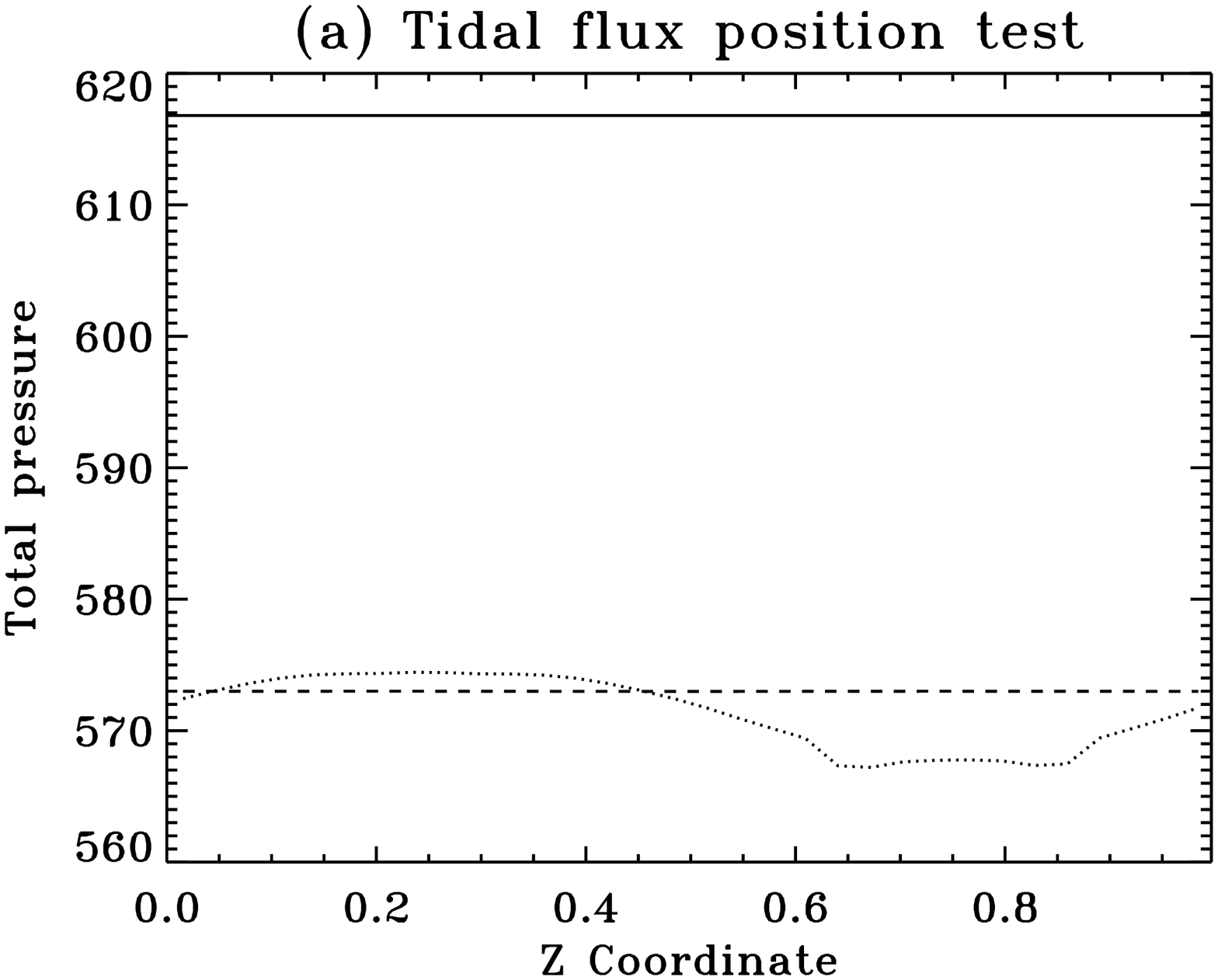,width=6cm,angle=+90}&
\psfig{file=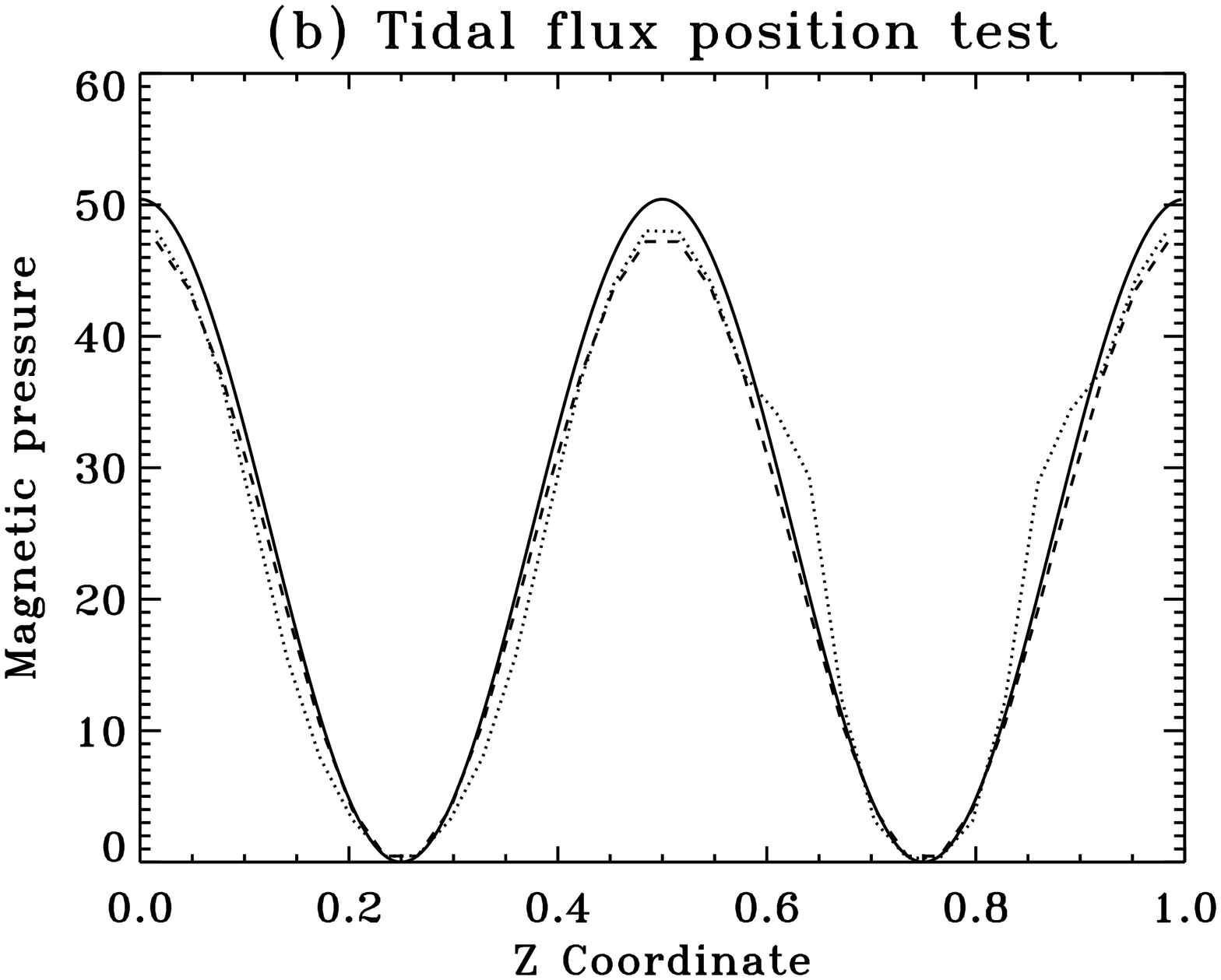,width=6cm,angle=+90}\\
\end{tabular}
}
 \caption{A benchmark with shear, resistivity, viscosity and heating
   (see section \ref{all}).  The wave number is $k=2\pi$ and $\rho=1$.
   We display the total pressure (a, left panel) and the magnetic
   pressure $(b_x^2+ b_y^2)/2$ (b, right panel). The time is $t=6$ which
   corresponds to 12 e-folding times ($s=1/2$).  The {\it solid line}
   is the analytical solution, the {\it dashed line} is for a total
   energy scheme where the flux of tidal potential $\rho \Phi \vv$ is
   computed jointly with the transport step and the {\it dotted line}
   is a total energy scheme where this flux is computed with the tidal
   source term. 
}
\label{mri}
\end{figure}

\section{Numerical viscosity and resistivity}

  In appendix \ref{method}, we present a method to estimate the
numerical resistivity and viscosity in a code.  The idea is to probe
the numerical dissipation in the absence of explicit resistivity and
viscosity, and to determine the effective numerical dissipation coefficients
by fitting results to our analytical solutions that include viscosity
and resistivity. In the following, we write for short $\etan$ and
$\nun$ for the numerical resistivity and viscosity. As explained in
the appendix \ref{method}, we measure directly $(\etan+\nun)k^2$ and
$(\etan-\nun)k^2$ respectively to second and first order in
$(\etan-\nun)k/v_A$. We then deduce the values for $\etan$ and $\nun$.
  We used this method mainly on the internal energy scheme version of
the Zeus3D code since it is the version that is generally used in
published applications.    


\subsection{General trends}
\label{trends}
  Our method provides a direct estimate for the numerical dissipation
in a code.  Therefore it gives the numerical floor for the physical
viscosity and resistivity in a given code.  For codes devoid of a
viscous or resistive term, it also allows to compute the effective
Reynolds and Prandtl numbers.  We now investigate general trends of
the numerical dissipation.

  We first examine wave numbers along the vertical direction.  In
figure \ref{depend}, we examine the dependence of $\etan$ and $\nun$
with various parameters. The Courant number (or Courant coefficient)
is a parameter that controls the time step of a code. In the Zeus3D
code it is defined as \beq C=\frac{\Delta t}{\Delta x}\max
\left(\sqrt{v^2+c^2+v_A^2} \right) \eeq where $v$, $c$ and $v_A$ are
the local speed, sound speed and Alfvén speed in the fluid, $\Delta x$
is the size of a zone, $\Delta t$ is the size of a time step and the
maximum is taken over all grid zones.  We measured the dependence on
resolution, wave number, perturbation amplitude and mean field
amplitude for three different Courant numbers:  0.01, 0.1 and 0.5.
 we display the results only for a Courant number of 0.1  and we
discuss the differences when applicable. We first ran a standard run
with parameters $\beta=2/B_0^2=400$, $k=2 \pi$, an amplitude of $|\de
b_x|=0.001$, $\rho=1$, a Courant coefficient of $C=0.1$ and a spatial
resolution of 32 zones in all three directions (hence $\Delta x=1/32$
since we use a physical length of 1 for the size of the box).  We then
varied each parameter in turn away from these values.

Figure \ref{depend}a and \ref{depend}b show that $\etan$ and $\nun$
scale linearly with the size of the time step and as the square of the
size of a grid cell.  An interpretation of these trends is that our
scheme is 1st order in time but 2nd order in space.  Note that
at a Courant coefficient of $0.5$, the numerical $\etan$ changes
sign. As a whole, the numerical scheme remains stable in the sense
that $\etan+\nun$ is always positive. However, $\etan$ or $\nun$
individually could be negative.  $\etan<0$ indicates that the MHD part
of the time step behaves like antidiffusion.
 Antidiffusivity in Zeus was already noted by \cite{F02} who also pointed out
that lower Courant numbers lower antidiffusion.  More recently,
\cite{F07I} also pointed out antidiffusion in Zeus at large
scales. Here, we  quantify the effect in more detail.
  The wave number with
the lowest numerical resistivity turns out to be
$\gb{k}=2\pi(\hx+\hy+\hz)$. The resistivity of this mode is negative
for all Courant coefficients above 0.12 (see dashed line on figure
\ref{depend}b). Such negative values for the resistivity are only
found for wave numbers with coordinates lower or equal than 2: only
the largest scales are affected. With the Zeus3D code, it might
nevertheless be safer to adopt Courant coefficients below 0.5 or to
include some minimal amount of physical resistivity in the code. 
Including physical dissipation has the advantage that it will also
improve the energy budget, as noted in the previous section.

In figure (\ref{depend}b) it appears that the dissipation has a finite
limit as the time step tends toward zero. Indeed, the finite space
resolution does not allow the scheme to achieve an infinite precision.
Similarly, in figure (\ref{depend}a) the scaling of the numerical
dissipation is a power of -2 in the number of zones at low space
resolution, but at high space resolution it turns into a shallower
power of -1.  Indeed, the Courant number is kept fixed (hence $\Delta
t/\Delta x$ is fixed) and the scheme is second order in space but only
first order in time: at high resolution, the numerical dissipation is
dominated by the order of the time integration scheme. At a higher
Courant number of 0.5, the shallower slope of -1 occurs at even lower
space resolution.  At a Courant number of 0.01, the slope of -2 is
seen over the whole range of space resolution we tested.  Note that
the numerical resistivity also turns out to be negative for the highest
resolutions at Courant numbers 0.1 and 0.5.

Figure \ref{depend}c shows the dependence of $\etan$ and $\nun$ on the
wave number.  Unlike a physical viscosity or resistivity, $\etan$ and
$\nun$ vary according to the length scale, with a maximum at
$k=8\times 2\pi\hz$ (four grid points inside each
wavelength).  There is no clear scaling but figure \ref{depend}c
suggests that the total dissipation behaves roughly like a power with
an exponent between 1 and 2 (1.6 seems to be the best match) until it
reaches the maximum dissipation. The quantity $(\etan-\nun)k/v_A$ can be as
high as $0.5$ for $k=8\times 2\pi\hz$, so the method of appendix
\ref{method} is not accurate for higher wave numbers.

  Our estimates show that numerical dissipation is nearly independent
of the amplitude of the initial perturbation (although there is a
slight dependence on it at a Courant number of 0.5). However, we
observe a strong dependence on the amplitude of the mean magnetic
field (see figure \ref{depend}d). The numerical viscosity appears to
be directly proportional to the mean magnetic field at low Courant
number (0.01) with some additional dissipation at low $\beta$ for
higher Courant numbers. The numerical resistivity obeys the same law,
with an additional change of sign at low values of $\beta$ (note: it
remains positive at low $\beta$ for a Courant number of 0.01). This
explains why the benchmarks of the previous section (which are for
large mean field amplitudes) have strong dissipation compared to the
standard runs of the present section. This might be an issue for the
computation of the saturated state of the MRI with Zeus3D. If
numerical dissipation does increase with the turbulent magnetic energy,
this could affect the total effective dissipation in the system.

  To summarise these results, we suggest to approximate the total
numerical dissipation with the following scaling formula: 

\beq
\label{dissip}
\etan+\nun \simeq 0.76 \Delta x^2\beta^{-\frac12}\left(\frac{k}{2 \pi}\right)^{1.6}
+1.08 \Delta x C \beta^{-1} 
\mbox{.}
\eeq 
  We calibrated both coefficients of this formula on figure
\ref{depend}b and the exponents for $\Delta x$, $C$, $k$ and
$\beta$ are obtained from figures \ref{depend}a, \ref{depend}b,
\ref{depend}c and \ref{depend}d respectively. Formula (\ref{dissip})
should therefore be taken only as indicative for values of parameters
not too far from those tested here. Furthermore, as
stated, the scaling in $k$ should also be taken with caution (see
figure \ref{depend}c).

\begin{figure}
\centerline{
\begin{tabular}{cc}
\psfig{file=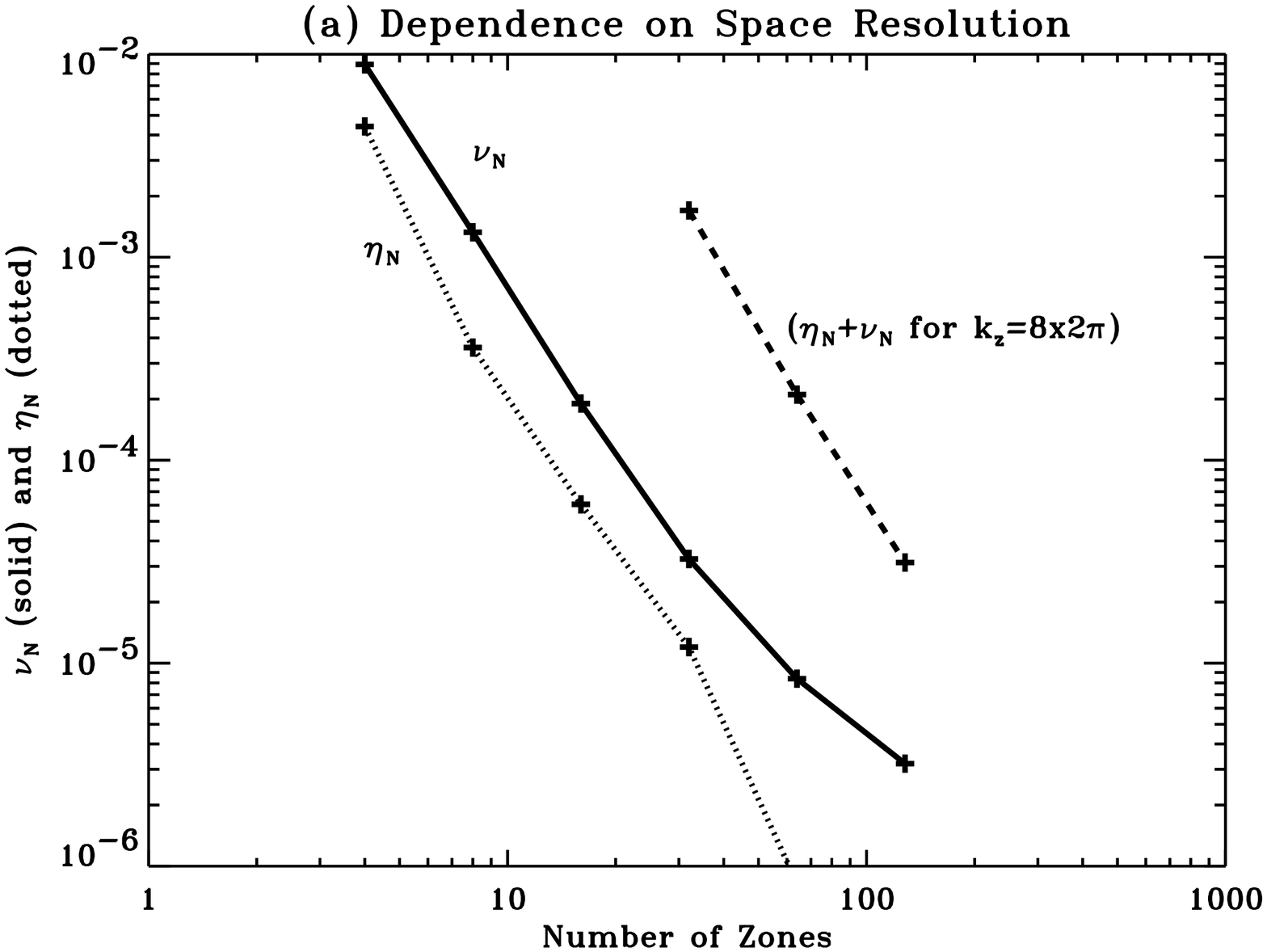,width=6cm,angle=+90}&
\psfig{file=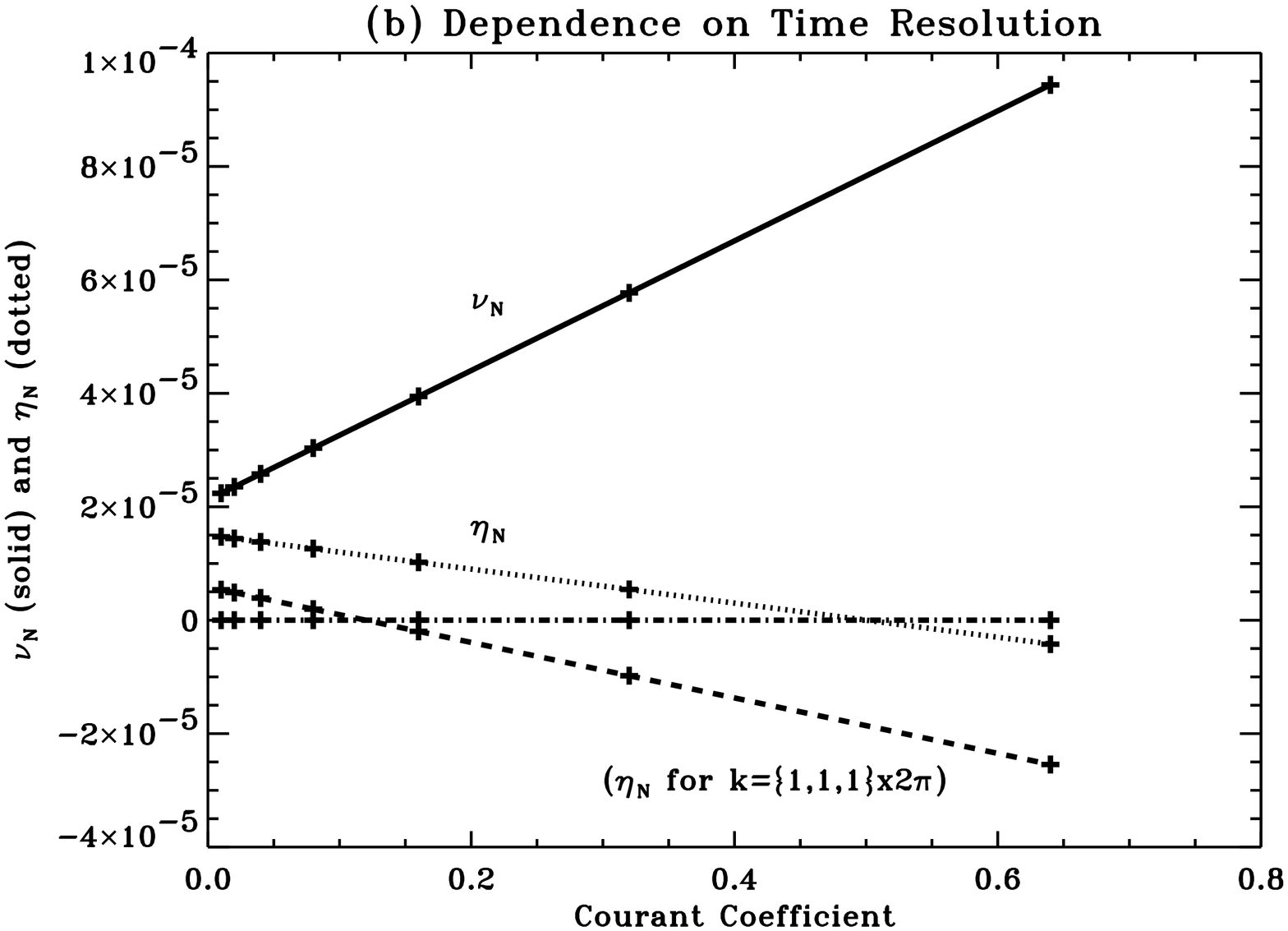,width=6cm,angle=+90}\\
\psfig{file=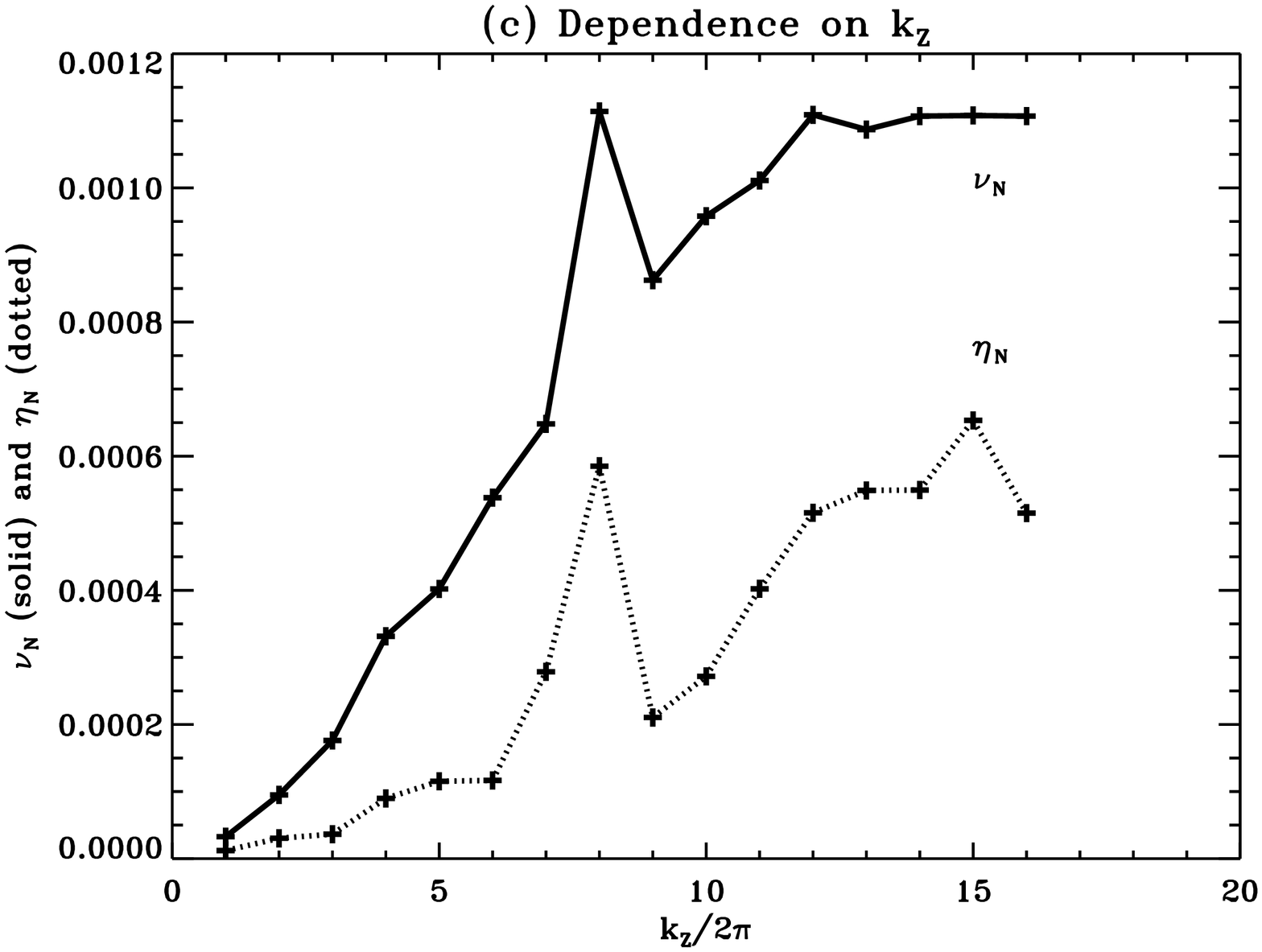,width=6cm,angle=+90}&
\psfig{file=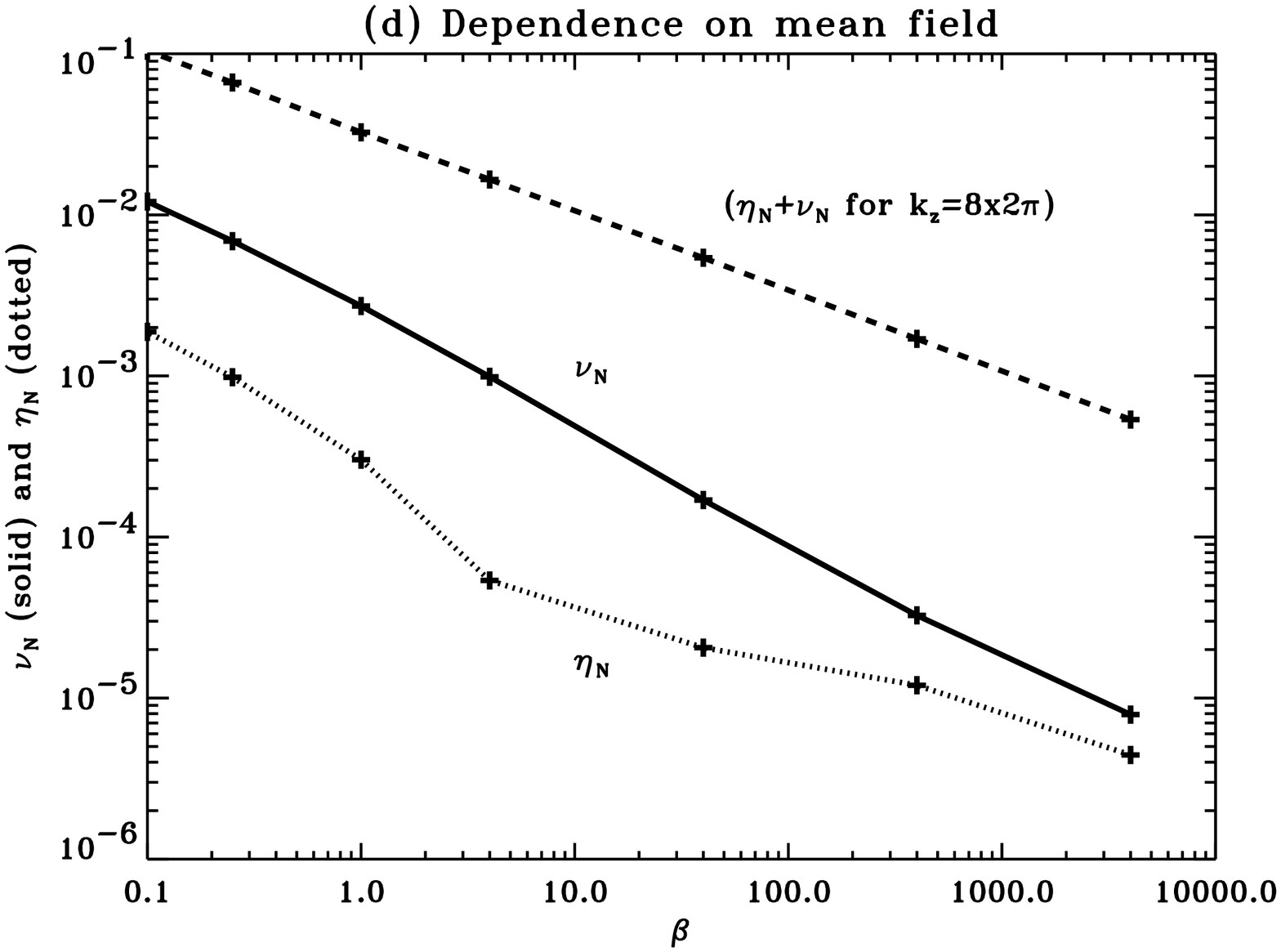,width=6cm,angle=+90}\\
\end{tabular}
} 
 \caption{We plot the numerical viscosity $\nun$ (solid) and
resistivity $\etan$ (dotted) vs various parameters such as (a) total
zone number (the dashed line is $\etan+\nun$ for $\gb{k}=2\pi\times 8
\hz $), (b) Courant coefficient \citep[see][]{ZI,ZII}(the dashed line
is $\etan$ for the wave vector $\gb{k}=2\pi(\hx+\hy+\hz)$ and the
dash-dotted line indicates the zero threshold.), (c) $k_z$ for a
mode with $\gb{k}$ along $\hz$ and (d) amplitude of the mean field (the dashed line is
$\etan+\nun$ for $\gb{k}=2\pi\times 8 \hz $) }
\label{depend}
\end{figure}

\subsection{Anisotropy}

  Our method allows us to quantify the anisotropy of the
numerical dissipation.  We measured the numerical dissipation for all
wave vectors in the Fourier domain of the box with coordinates of the
form $k_i=2\pi n$ with $i=x,y,z$ and $0 \leq n \leq 16$ (a grid of
$17^3-1$ measurements).  Many shearing box simulations actually use
half the resolution in the azimuthal $y$ direction compared to the
radial $x$ and vertical $z$ directions.  We therefore did the same
measurements (with $k_y \leq 8 \times 2\pi$) on a cubic box with
32x16x32 zones in which the grid cells have an aspect ratio 1:2:1.

\subsubsection{Cubic grid cells}
   In figure \ref{anis}a we plot all $\etan+\nun$ measurements for the
$32^3$ (cubic cells) simulation against the norm of the wave vector.
The overall shape of this diagram roughly follows figure \ref{depend}c
with a maximum of dissipation at 15$\times 2\pi$. Even for cubic grid
cells, the numerical dissipation already shows some degree of
anisotropy: at a given wave number it varies widely.  We detail the
distribution of this spread in figure \ref{grid}a for wavenumbers $k$
which have $2\pi N_k\leq k < 2\pi (N_k+1)$ with $N_k=15$. Wave vectors
with the highest dissipation are those that point towards a cartesian
axis.  For a fixed $|k|$, wave vectors along an axis maximise the size
of a single component.  We therefore suggest that the numerical
dissipation at a given wave vector is dominated by the dissipation at
its maximum coordinate.  For wave vectors of norm $|k|$ higher than
16, the three coordinates have similar values, hence the numerical
dissipation is more and more isotropic.  Interestingly, the numerical Prandtl
number $$Pm_{\rm N}=\nun/\etan$$ is quite isotropic for all wave numbers and
slightly decreases from 2 at small wave numbers to 1 at large wave numbers (see
figure \ref{anis}c; the very small wave numbers have higher Prandtl numbers,
but the numerical dissipation is much lower there). The isotropy of
the Prandtl number is even better at lower Courant coefficients
($C=0.01$), with a mean value closer to (slightly above) 1 and a spread
between 1 and 2. It is interesting to compare these results to the recent
work of \cite{F07II} who estimate Prandtl numbers between 2 and 4 for Zeus.
It is also striking that all our measured Prandtl numbers are greater than 1.

  As  mentioned, a few directions yield a negative resistivity.
The corresponding wave vectors at $C=0.5$
have their coordinates amongst the following list: (1,1,1), (1,1,0),
(2,1,1), (2,2,1), (2,2,2) and their permutations.


\subsubsection{Elongated grid cells}

  For elongated cells, the diagram \ref{anis}b is only slightly more
complicated.  It is similar to figure \ref{anis}a, but replicates its
pattern extended by a factor 2 in amplitude and squeezed by a factor 2
in wave numbers.  This additional feature results from the halved
resolution in the $y$ direction.  As shown on figure \ref{grid}b, the
dissipation is not symmetric to $x$-$y$ exchange.  On the contrary,
$y$ wave vectors undergo much larger dissipation.  This shows up even
more at smaller wave numbers as seen in figure \ref{grid}c and
\ref{grid}d.  However, the Prandtl number does not show more
anisotropy than in the case of a cubic cell: numerical resistivity and
viscosity react in the same way to the resolution loss in the $y$
direction.
  

\begin{figure}
\centerline{
\begin{tabular}{cc}
\psfig{file=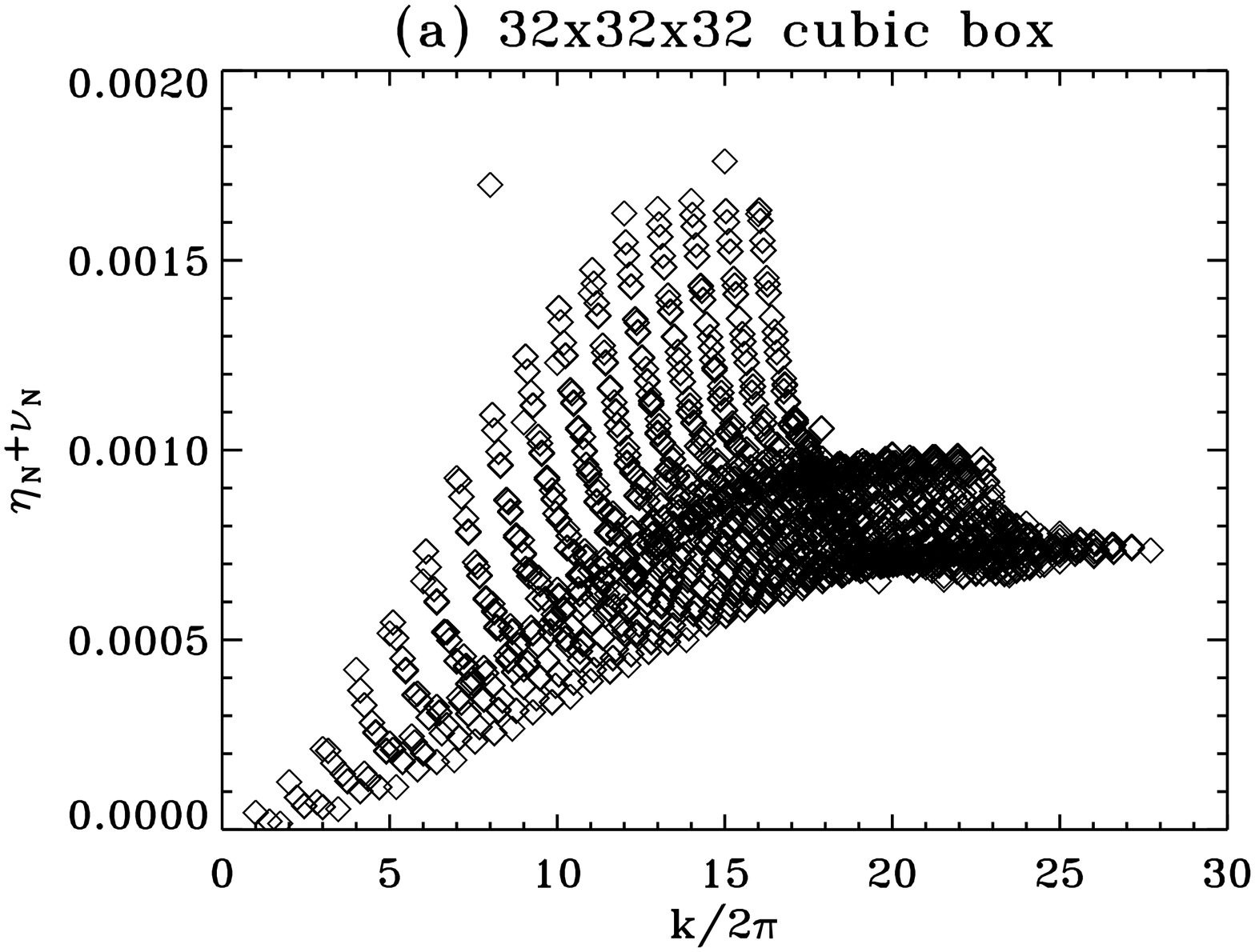,width=6cm,angle=+90}&
\psfig{file=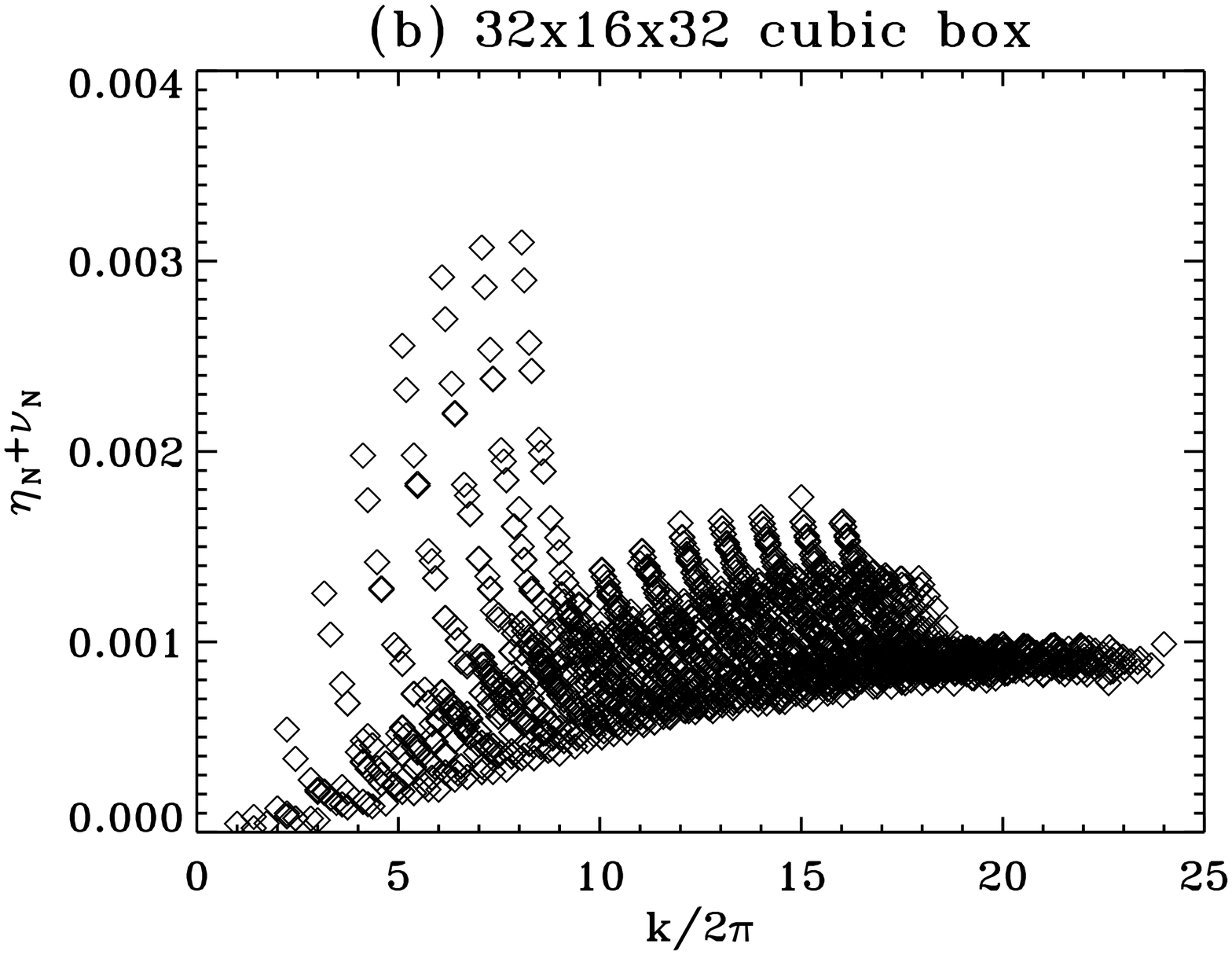,width=6cm,angle=+90}\\
\psfig{file=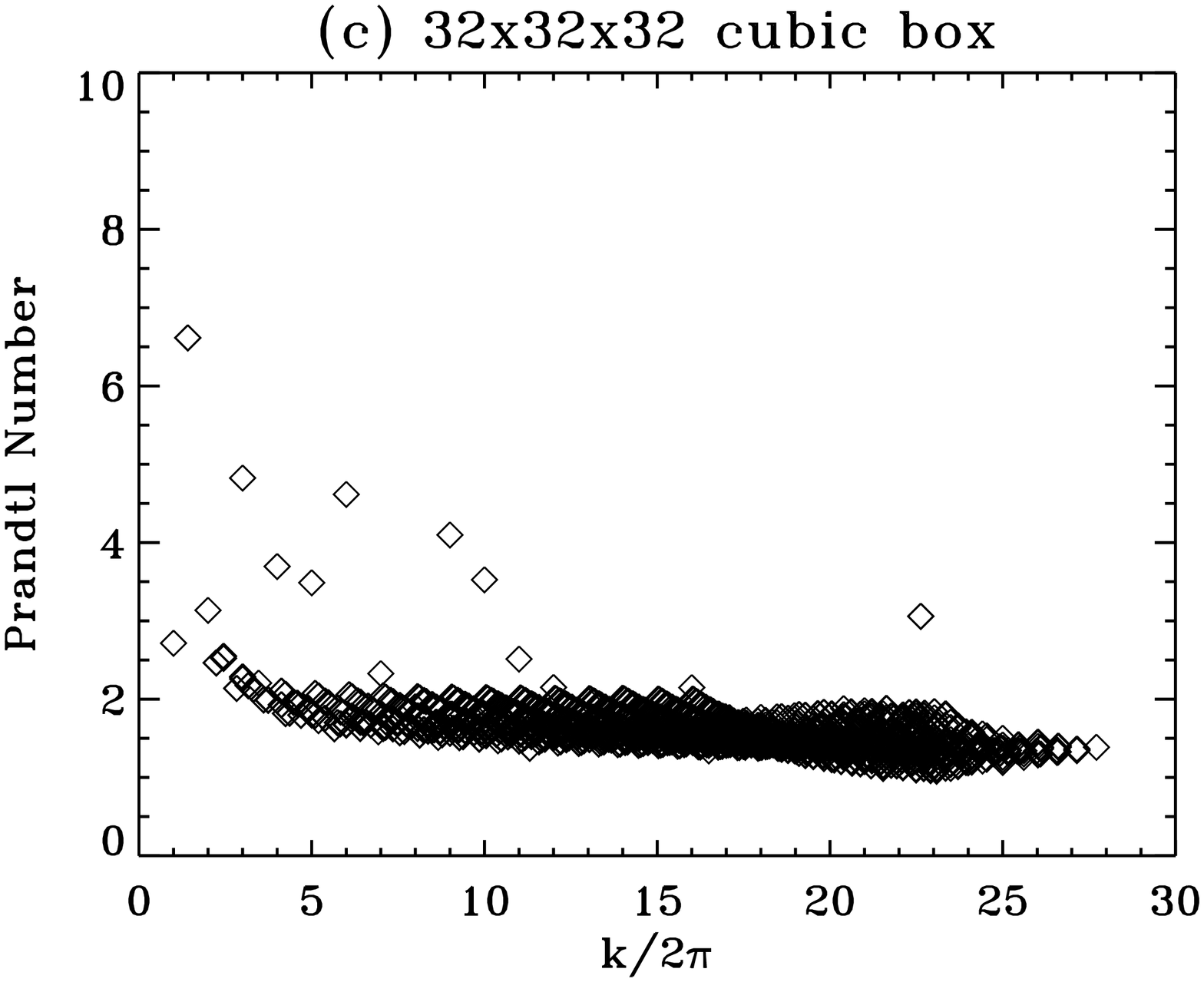,width=6cm,angle=+90}&
\psfig{file=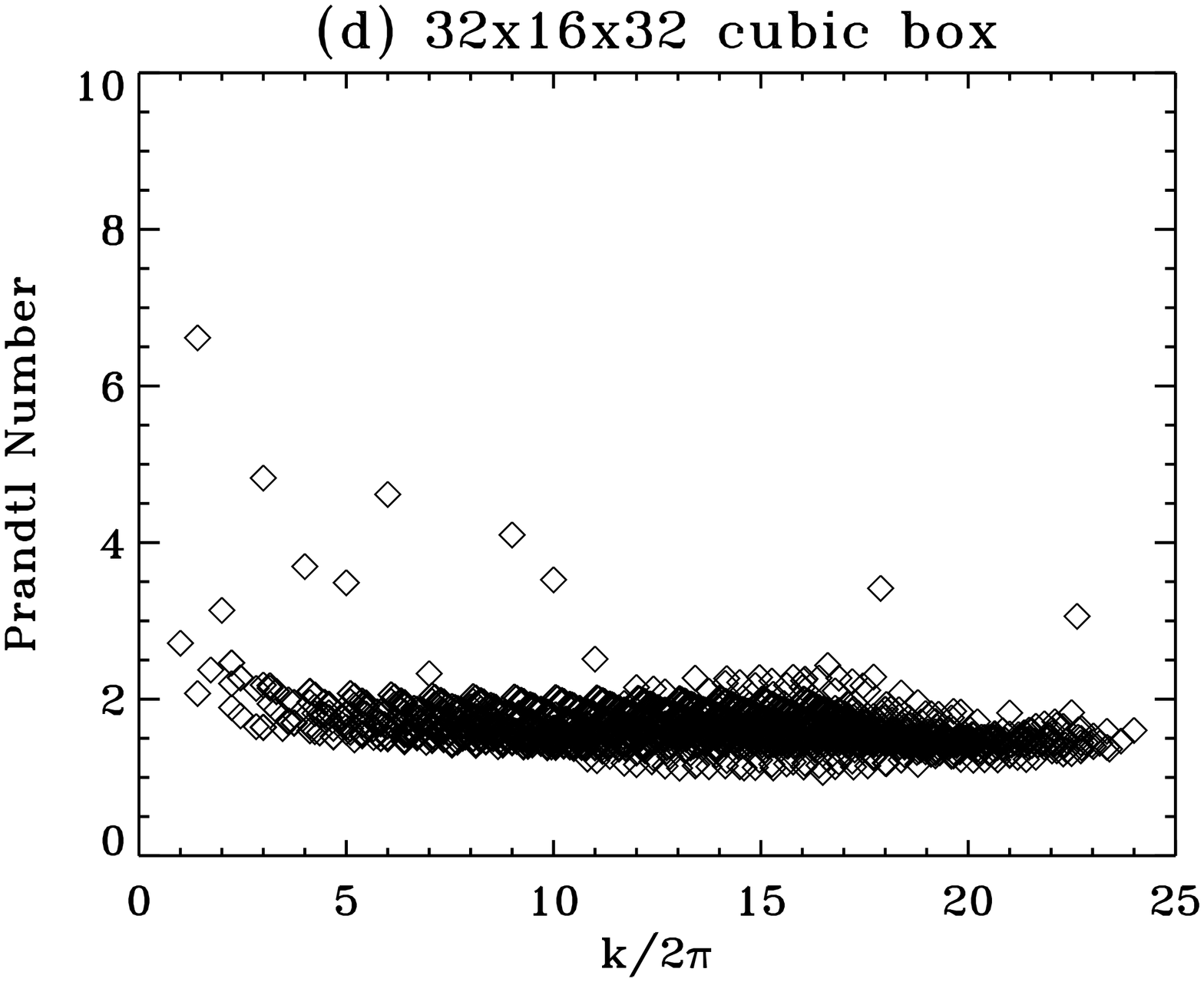,width=6cm,angle=+90}\\
\end{tabular}
} 
\caption{We show the total numerical dissipation $\etan+\nun$ (panels
a and b, upper side) and numerical Prandtl number $Pm_{\rm
N}=\nun/\etan$ (panels c and d, lower side) vs.  $k$ for every
wavevector of the computational box.  Panels a and c (left hand side)
are for a 32x32x32 zones computational box.  Panels b and d (right
hand side) are for a 32x16x32 zones box with elongated cells.  }
\label{anis}
\end{figure}

\begin{figure}
\centerline{
\begin{tabular}{cc}
\psfig{file=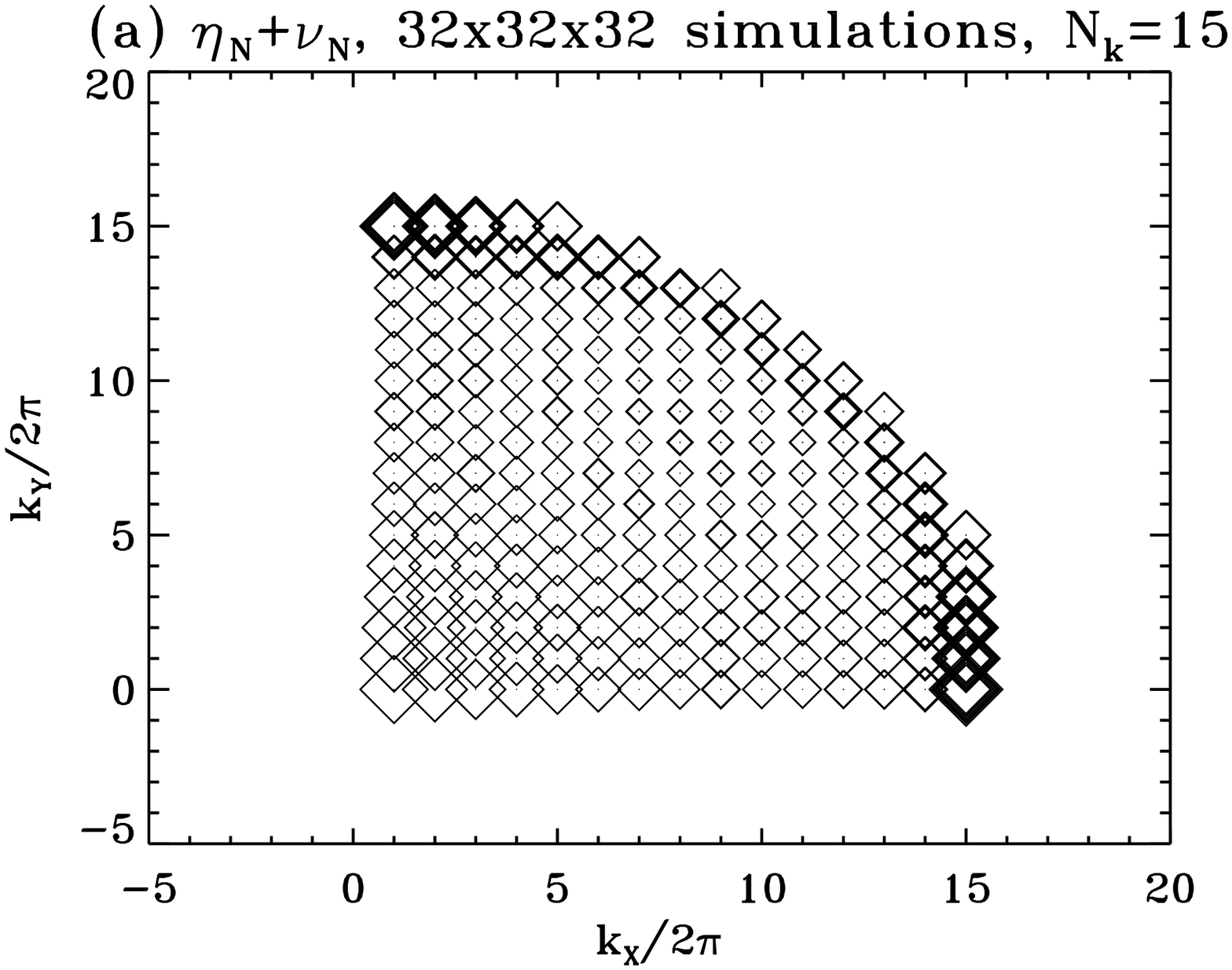,width=6cm,angle=+90}&
\psfig{file=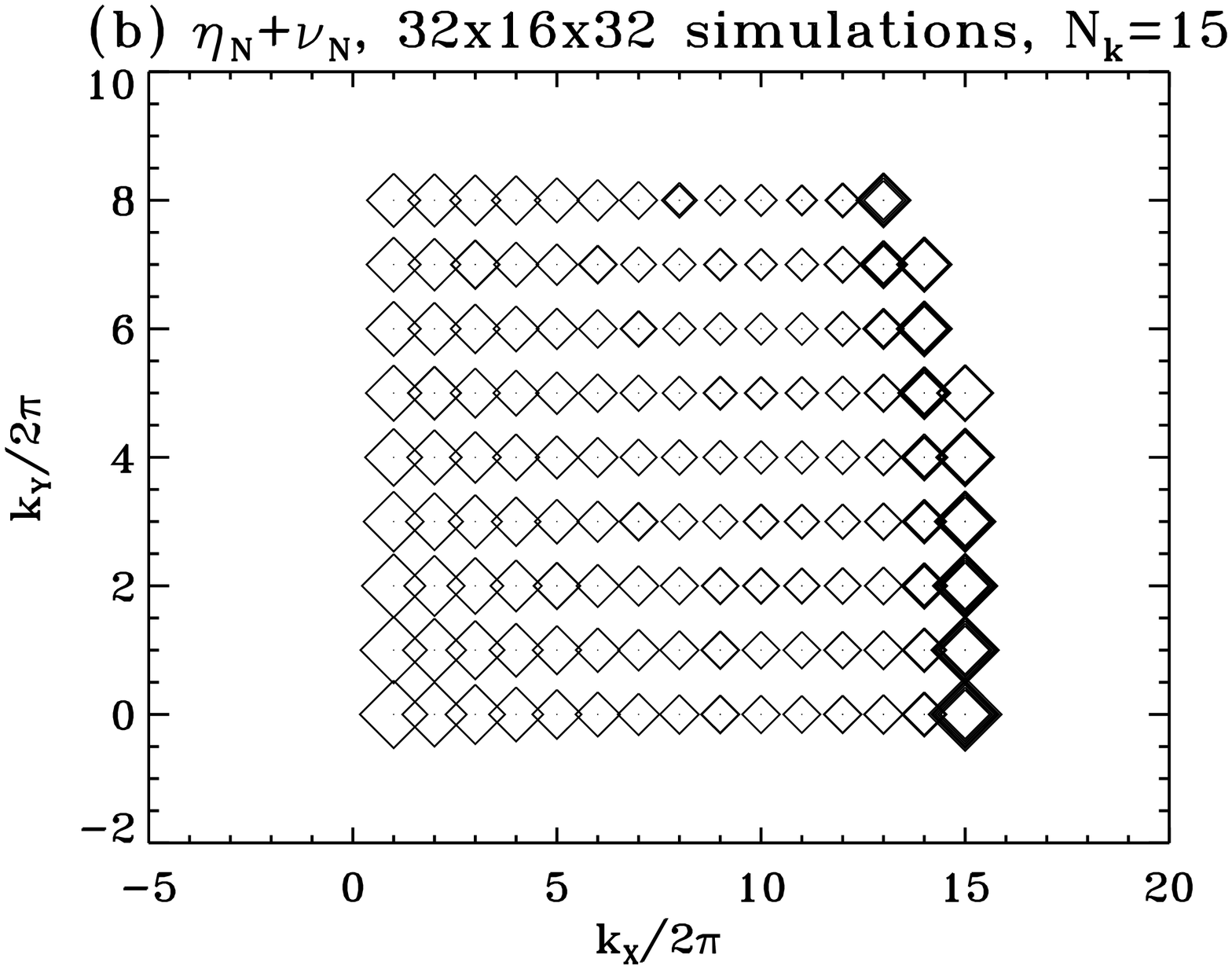,width=6cm,angle=+90}\\
\psfig{file=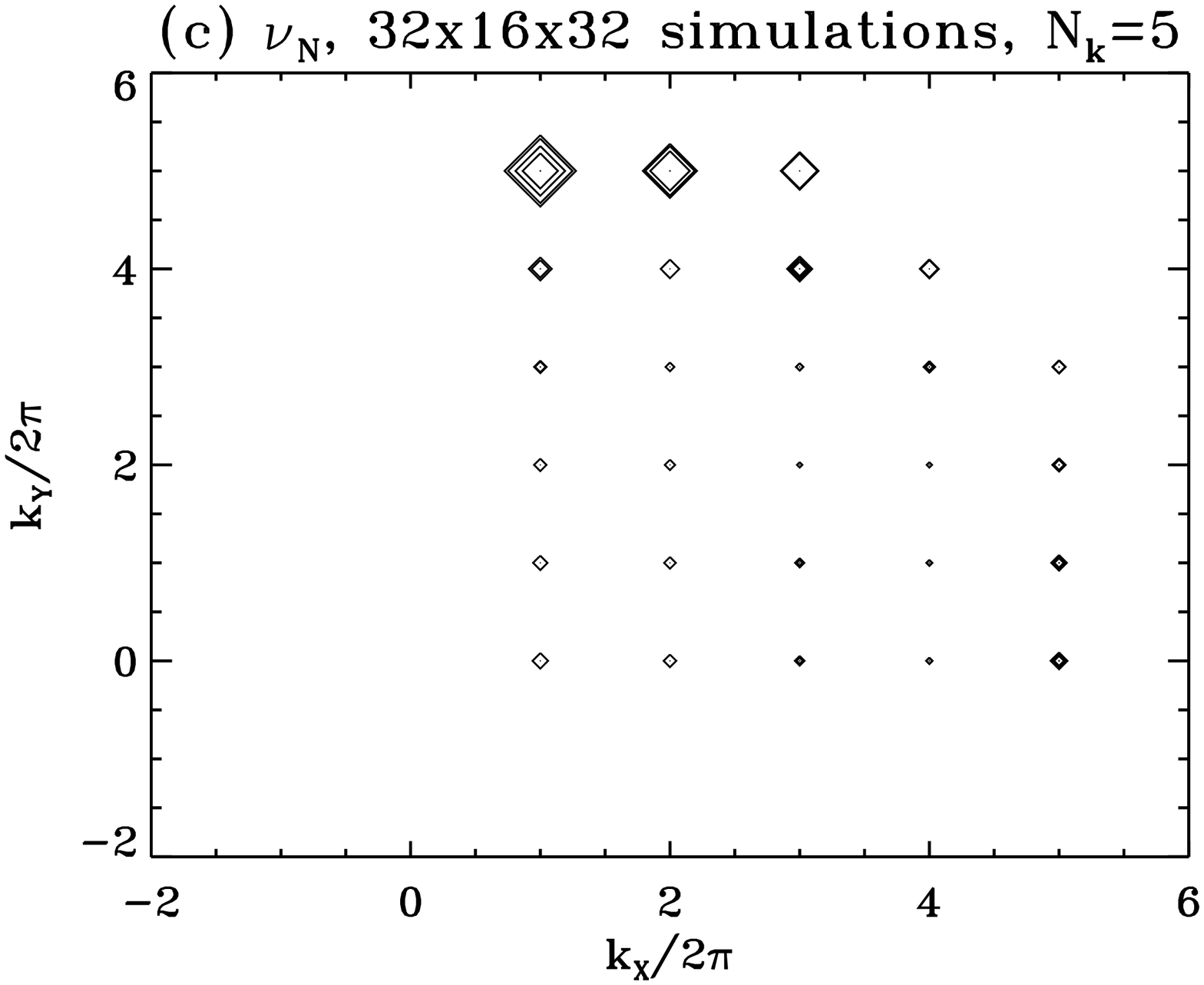,width=6cm,angle=+90}&
\psfig{file=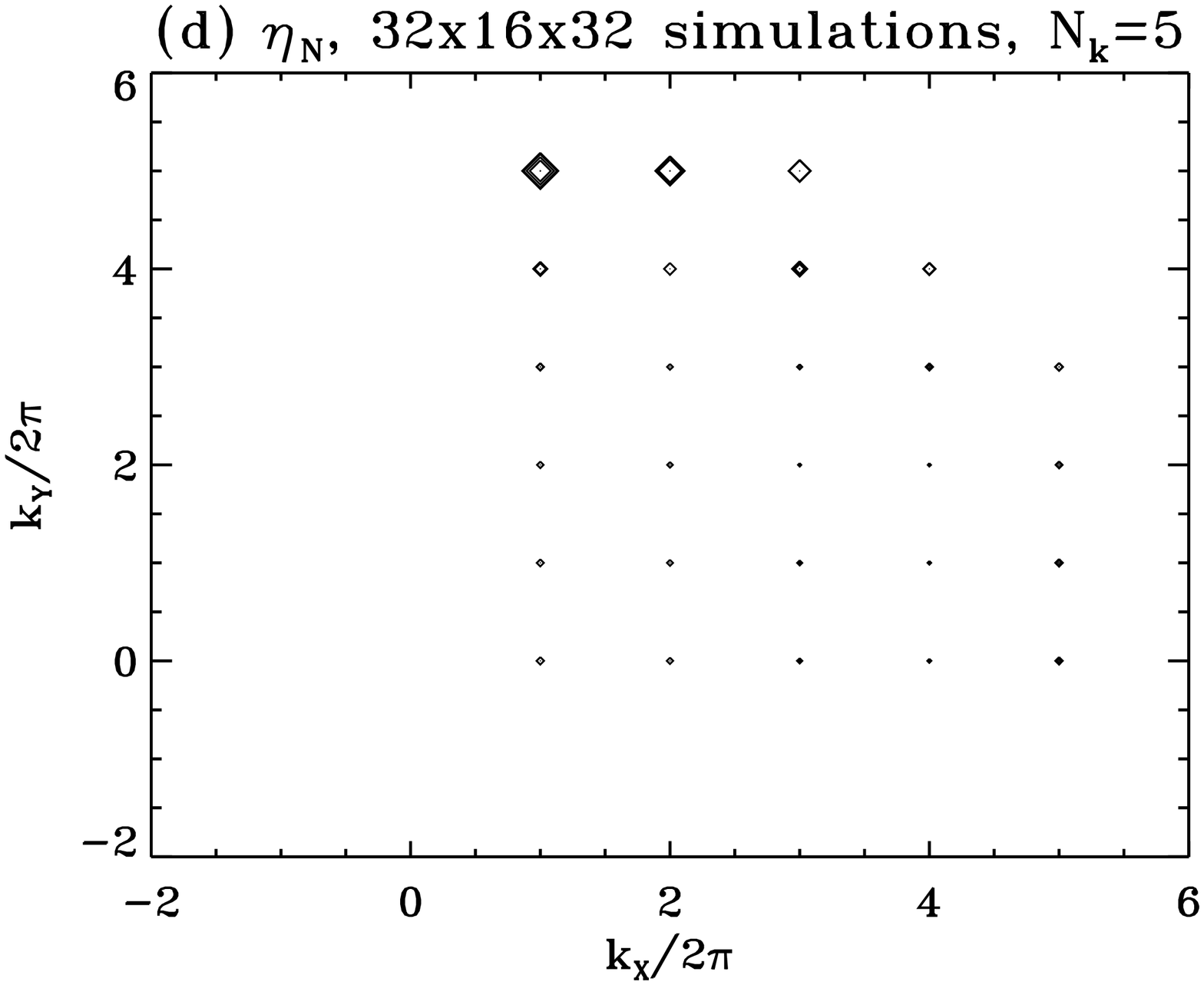,width=6cm,angle=+90}\\
\end{tabular}
} 
 \caption{We show the total numerical dissipation $\etan+\nun$ (panels
a and b, upper side) and numerical resistivity $\nun$ (panels c and d,
lower side) for each wave number with positive coordinates and such
that $2\pi N_k\leq k < 2\pi (N_k+1)$.  The size of the symbols codes
for the magnitude of the quantity plotted and their position marks
their $x$ and $y$ coordinates.   Results for wave vectors with same $x$
and $y$ coordinates but differing $z$ coordinates are overplotted on
top of each other.   Panels (a) and (b) compare the simulations with
cubic cells (a) to the simulation with elongated cells (b) in the case
where $N_k$=15.    Panels (c) and (d) compare numerical viscosity (c)
and resistivity (d) in the case with elongated grid cells for
$N_k=5$.  }
\label{grid}
\end{figure}

\subsection{Scheme}
\label{scheme}

  We tested our code in various configurations to 
investigate the impact on numerical dissipation.  We found that 
isothermal simulations are slightly less dissipative than adiabatic
simulations (with numerical resistivity more negative in general).  We
could hardly see any difference between the internal and the total
energy schemes.  We also found that to use the non-linear artificial
resistivity as coded in \cite{ZI} did not change the numerical
dissipation: our tests did not trigger significant artificial
viscosity because of the incompressible nature of our test flows.

\section{Discussion}

 \subsection{Incompressibility}
  We discuss here a few caveats, limitations and possible extensions
of our method.   First, we wish to stress that the condition of
incompressibility on the modes is a crucial one.   Indeed, any change in
density will alter the $1/\rho$ factor in the Euler equation,
introducing additional non-linearities that might be difficult to
address with analytical tools.  

  In section 7 we have measured an
equivalent numerical viscosity for {\it torsional Alfv\'en waves}
only.  This gives a first estimate of the numerical dissipation, but
an arbitrary MHD flow cannot be decomposed into such modes.  For example
we cannot probe any viscosity associated to compressible flows  (see section
\ref{scheme}).  On the other hand, the viscosity of compressible flows
could be measured by studying the width of shock fronts, or with
damped magnetosonic  waves in the linear regime.

  
  \subsection{Boundary conditions}
  Periodic boundary
conditions in $z$ are essential for our analysis.   For example,
reflective boundary conditions will mix two Fourier modes which very
likely will open the way for a cascade at many other wave
numbers.    

  It should be noted that our analytic shear solutions do not strongly
test the implementation of shearing box boundary conditions.  Indeed,
except for the mean steady flow, our solutions for non-zero $\Omega$
depend {\it only} on the $z$ coordinate.  For example, these
benchmarks could not tell if the code uses periodic boundary
conditions in the $x$ direction or shearing box boundary
conditions.
  
  To improve this, we could need to find solutions
with spatial variation in more than one direction.  A non-zero $k_x$
is in fact perfectly tractable, and the equations hardly change if one
uses the expression $(\gb{k.B_0})^2/\rho=k_z^2 B_0^2/\rho$ instead of
$\ww$.  However, it will not probe more efficiently the shearing box
boundary conditions: periodicity in $x$ would still be
indistinguishable from shearing box conditions for most variables.

In order to probe the shearing box boundary conditions, a non-zero
azimuthal wave number is needed.  Unfortunately, a non-zero $k_y$
yields an Eulerian wave vector changing in time \citep{MRI4}. In that
case, the homogeneity condition changes in time and the total pressure
gradient cannot be dealt with at all times. However, it is worth noting that 
 the total pressure term is still in this case the only non-linear
term. Semi-analytic solutions of the equations without the pressure term
can be found for MHD shearing waves. We plan in future work to 
benchmark the MHD shearing box boundary conditions by using such equations.


  \subsection{Thermal diffusion}
 A thermal diffusion coefficient can very easily be included in our
analytical solutions.  Under the assumption of a uniform density, the
thermal diffusion term in equation (\ref{dfta}) is proportional to
\beq \chi \triangle p=\chi \triangle \left(p_{\rm
tot}-\frac12\Re[\Bb]\bdot\Re[\Bb]\right) =\chi 2k^2
\Re[\Bb]\bdot\Re[\Bb] \eeq where $\chi$ is the uniform thermal
diffusion coefficient. We recall that the term due to cooling in
equation (\ref{dfta}) is $\alpha/2~\Re[\Bb]\bdot\Re[\Bb]$. To include
thermal diffusion in our formalism is hence equivalent to use
$\alpha+4k^2\chi$ in place of $\alpha$.

\subsection{Total pressure gradient}
The assumption of a homogeneous total pressure
 is the cornerstone of our analysis.   This requirement might not
be as strong as it seems at first glance.   Indeed, the gradient of
total pressure in the Euler equations naturally drives MHD flows
towards a state of uniform total pressure.   As a result, our solutions
should be close approximations to the exact MHD flows even in cases
when the condition of homogeneity is not met, provided that the real flow
remains at a nearly constant density.   

\section{Conclusions}

  This paper consists of variations on a theme: the channel solution.  
We have extended previously known analytical solutions to more general and
more physical cases, including viscosity, resistivity and cooling.   We also
showed the connection between torsional Alfv\'en waves and channel solutions.  

  We used these solutions to calibrate the implementation of a
conservative scheme in Zeus3D.  We also measured the numerical
resistivity and viscosity of torsional Alfv\'en waves in Zeus3D.  In
particular, we showed that lower time steps should be used in Zeus3D
in order to  guarantee a positive resistivity when no physical resistivity
is used. We would rather recommend to use a minimal amount of physical
resistivity.  It also is best to use isotropic resolution since the
numerical dissipation is more anisotropic for elongated cells. Finally 
we find a  dependence of the numerical dissipation on the amplitude
of the magnetic field.
  
  Although in this paper we stressed the numerical applications of
these solutions, they are of interest in their own right.  In
particular, we have established a stronger basis for understanding the
stability analysis of channel solutions: it should now be possible to
compute parasitic instabilities with improved microphysics.  As a
result, we hope to better understand the saturation properties of MRI
turbulence.

\section*{Acknowledgements}

Many thanks to S. Fromang for providing us with his version of the
Zeus3D code. We thank the anonymous referee for a thorough report
which significantly improved the quality of the paper.  This work was
supported by a Chaire d'Excellence awarded by the French ministry of
Higher Education to S. Balbus.

 \appendix

\section{Propagating disturbances with zero shear}
\label{ccomp}
  We work here in the framework and notations of section
\ref{nonpolar}.  We first assume that there exists a common complex
root $s$ to $R$ and $Q$.  Then its complex conjugate $\bar{s}$ also is
a common root.  Being only second degree, $R$ and $Q$ have the same
roots $s$ and $\bar{s}$. Hence they differ only by a real
proportionality constant, and the remainder $R_1$ of the Euclidian
division of $R$ by $Q$ needs to be identically zero:

\beq
\label{R1}
R_1(s)=a_1 s + a_0
\eeq
with
\beq
a_1=\nu-\eta-\frac32\,\frac{\ww}{\nu}
\eeq
and
\beq
a_0=(\nu-\eta)\eta+\ww\left(1+\frac{-\alpha+2\eta}{2\nu}\right)
\mbox{.}
\eeq

Both coefficients $a_0$ and $a_1$ must vanish if $R$ and $Q$ are to have
a common complex root.
This puts two constraints on the three remaining parameters $\eta$, $\nu$
and $\w$ so that $\eta$ and $\nu$ can be expressed in terms of $\w$. 
Setting $a_1$ equal to zero, we get
\beq
\label{equeta}
\eta=\nu-\frac{3\ww}{2\nu}
\mbox{.}
\eeq
  We now use this expression into the equation $a_0=0$  which
yields a quadratic equation for $\nu$:
\beq
14\,\nu^2-2\ww\alpha \, \nu +15\, \ww=0
\eeq
which has only one real positive root 
\beq
\label{nu}
\nu=\frac1{14}\left(\alpha+\sqrt{\alpha^2+210\ww}\right)
\mbox{.}
\eeq
 Equation (\ref{equeta}) now provides the value for $\eta$
\beq
\label{eta}
\eta=\frac1{35}\left(6\alpha-\sqrt{\alpha^2+210\ww}\right)
\eeq
which is positive for $\alpha>\sqrt{6}|\w|$ .

  In order to get a common complex root to $P$ and $Q$, the
resistivity and viscosity need hence to be determined by expressions
(\ref{eta}) and (\ref{nu}).  Using these expressions for $\eta$ and
$\nu$ in the dispersion relation $R(s)=0$ provides the two actual
growth rates $s$ and $\bar{s}$ as

\beq
s_\pm =\frac{ -17\alpha-3\sqrt{\alpha^2+210\ww}}{140}
\pm i \sqrt{r}
\eeq
with
\beq
r=95 \ww+\alpha\sqrt{\alpha^2+210\ww}-\alpha^2
\mbox{.}
\eeq

  The corresponding solutions are therefore propagating disturbances (\ie
they have a non zero imaginary part) only when $r>0$ which is
equivalent to setting the condition $\alpha>-\frac{19}4|\w|$. Since we already
required the more stringent condition $\alpha>\sqrt{6}|\w|$ in order
to get $\eta>0$, physically plausible solutions exist only when
$\alpha>\sqrt{6}|\w|$ for propagating disturbances.

\section{Standing waves with zero shear}
\label{creal}
\label{standing}
  We work here in the framework and notations of section \ref{nonpolar}.
  We now assume that $s$ is a real common root to $P$ and $R$. In that
case, $R_1(s)=0$ immediately gives $s$ in terms of the parameters of the other
problem:

\beq
s=\frac{-2\nu\eta(\nu-\eta)+\alpha\ww-2(\eta+\nu)\ww}{2\nu(\nu-\eta)-3\ww}
\eeq Using $s$ in $P$ or $Q$, we finally arrive at a relational
constraint for the defining of the problem.  As an explicit example,
$\ww$ can be expressed in terms of $\alpha$, $\eta$ and $\nu$:
\beq
\ww=\frac1{18}\left[-\alpha^2+7\alpha(\eta+\nu)+2\nu^2-41\eta\nu-10\eta^2
+(-\alpha+2\nu+5\eta)
\sqrt{\alpha^2-\alpha(10\nu+4\eta)+\nu^2+44\eta\nu+4\eta^2}\right]
\mbox{.}
\eeq

 \section{A method to measure numerical resistivity and viscosity}
 \label{method}

  In section \ref{TAW} we showed that circularly polarised waves with
nonzero viscosity or resistivity are solution of the non-linear
equations. In principle, if we start our simulation with one of the
eigenmodes corresponding to the growth rate $s_\pm=\pm i \-w$
(equation \ref{growth} for $\eta=\nu=0$), we should obtain the time
evolution of a torsional Alfv\'en wave as a result of the computation.

  However, the finite grid and time stepping resolution introduce some
numerical defects. For example, figures \ref{alfven}a and
\ref{alfven}b show that numerical results undergo some dissipation. In
these figures, the dashed line corresponds to a wave with a slightly
lower amplitude than a pure torsional Alfv\'en wave after three
oscillation periods. This suggests that the numerical errors in the
code may behave like an equivalent viscosity and resistivity. In
principle we could define their effective values if we were able to
fit a model evolution to the actual numerical output of the code.

  In this appendix, we are motivated to compute the evolution of a system which
starts with the initial conditions for a torsional Alfv\'en wave (with
$\eta=\nu=0$), but which is evolved with some amount of viscosity
$\nun$ and resistivity $\etan$. Recall that $\eta=k^2 \etan $ and $\nu=k^2
\nun$ where $\etan$ and $\nun$ are the effective resistivity and
viscosity. We present the results to first order in
$(\eta-\nu)/\w=(\etan-\nun)k/v_A$.

  We choose the initial phase such that $\de b_x=1$. The initial conditions
for a torsional Alfv\'en wave give $\de u_x=1$ and $\de b_y=\de u_y=i$.

  We first assume that the code preserves well the initial uniform
density profile\footnote{We actually checked that to enforce $\rho=1$
in the code did not change much the measured $\nu$ and $\eta$.}.
According to section \ref{TAW} there exist only two incompressible
modes that can be excited with growth rates given by equation
(\ref{growth}). We decompose our initial conditions on the two
corresponding eigenmodes which have $\de
u_{x\pm}=(s_\pm+\eta)/(i\w)\,\de b_{x\pm}$:
\beq
\de u_x=1=\alpha_+ \frac{s_++\eta}{i\-w} + \alpha_- \frac{s_-+\eta}{i\-w}
\eeq
and
\beq
\de b_x=1=\alpha_++\alpha_-
\eeq
with 
\beq
\frac{s_\pm+\eta}{i\w}\simeq\frac{\eta-\nu}{2i\w} \pm 1
\eeq
and where $\alpha_+$ and $\alpha_-$ are the complex weights of the two eigen
modes.

We solve for $\alpha_+$ and $\alpha_-$ and retain the first order in
$(\eta-\nu)/\w$:

\beq
\alpha_+=
\frac12\left( 1+\frac{\w+i\frac{\eta-\nu}2}
{\sqrt{\ww-\left(\frac{\eta-\nu}2\right)^2}}\right)
\simeq 1+i\frac{\eta-\nu}{4\w}
\eeq
and
\beq
\alpha_-=1-\alpha_+ \simeq -i\frac{\eta-\nu}{4\w}
\mbox{.}
\eeq

The non-linear coupling between these two modes can only occur
through the total pressure gradient term and it happens that this term
vanishes to first order in $(\eta-\nu)/\w$. The temporal evolution of
the system can hence be approximated by its linear evolution

\beq
\de u_x=
\alpha_+ \frac{s_++\eta}{i\w} \exp(s_+t) 
+ \alpha_- \frac{s_-+\eta}{i\w}\exp(s_-t)
\eeq
and
\beq
\de b_x=\alpha_+ \exp(s_+t) +\alpha_- \exp(s_-t) 
\mbox{.}
\eeq

We finally recover the temporal evolution of the perturbed quantities as
\beq
\Re[u_x]=\exp\left(-\frac{\eta+\nu}2 t\right)
\left[ \cos(\w t+\gb{k.r})+\frac{\eta-\nu}{2\w}\sin(\w t)\cos(\gb{k.r})\right]
\eeq
and
\beq
\Re[b_x]=\exp\left(-\frac{\eta+\nu}2 t\right)
\left[ \cos(\w t+\gb{k.r})-\frac{\eta-\nu}{2\w}\sin(\w t)\cos(\gb{k.r})\right]
\mbox{.}
\eeq
The $y$ component of these fields can be recovered because of the circular
polarisation conditions $b_y=ib_x$ and $u_y=iu_x$.
We choose to recover $\eta$ and $\nu$ from their sum and difference
through the volumic averages of kinetic and magnetic energy:
\beq
\label{Ca}
\frac12<(\Re[\vu])^2+(\Re[\Bb])^2>=\exp\left(-(\eta+\nu) t\right)<\cos^2(\gb{k.r})>
\eeq
and
\beq
\label{Cb}
<(\Re[\vu])^2-(\Re[\Bb])^2>=(\eta-\nu)\,\frac{\sin(2\w t)}{2\w}\,<(\Re[\vu])^2+(\Re[\Bb])^2>
\mbox{.}
\eeq

Hence, we find the total dissipation $\eta+\nu$ from the exponential
decay of the kinetic plus magnetic energy. And we get the difference
$\eta-\nu$ from the relative difference between these two forms of
energy. Note that these final expressions yield $\eta+\nu$ with one
more order of accuracy in $(\eta-\nu)/\w$ than $\eta-\nu$.  We 
checked a posteriori that $(\eta-\nu)/\w$ is indeed small for all
measurements performed in this paper except for the highest wave numbers
(see section \ref{trends}).

  In order to save computing time, we evaluate $\eta$ and $\nu$ on the
very first time step of the simulation. Figures \ref{check}a and
\ref{check}b show that this provides reasonable estimates for
equations (\ref{Cb}) and (\ref{Ca}).  The match is in fact perfect
only for the first few time steps. The later discrepancy between our
model for the code diffusion and the actual results of the code is
probably due to the dispersive properties of the scheme which we do
not take into account.  This discrepancy actually narrows down at
higher Courant numbers which are known to be less dispersive.
However, figure C1 shows that our model captures the bulk of the
numerical artifacts.

Finally, in order to ensure that the initial fields have zero divergence, we
convert real wave numbers $k$ to discrete wave numbers
$k'=2\sin(k\Delta x/2)/\Delta x$ when we compute the relations between
the amplitudes of the fields. $\Delta x$ is the size of a pixel in the
units of the computation.

\begin{figure}
\centerline{
\begin{tabular}{cc}
\psfig{file=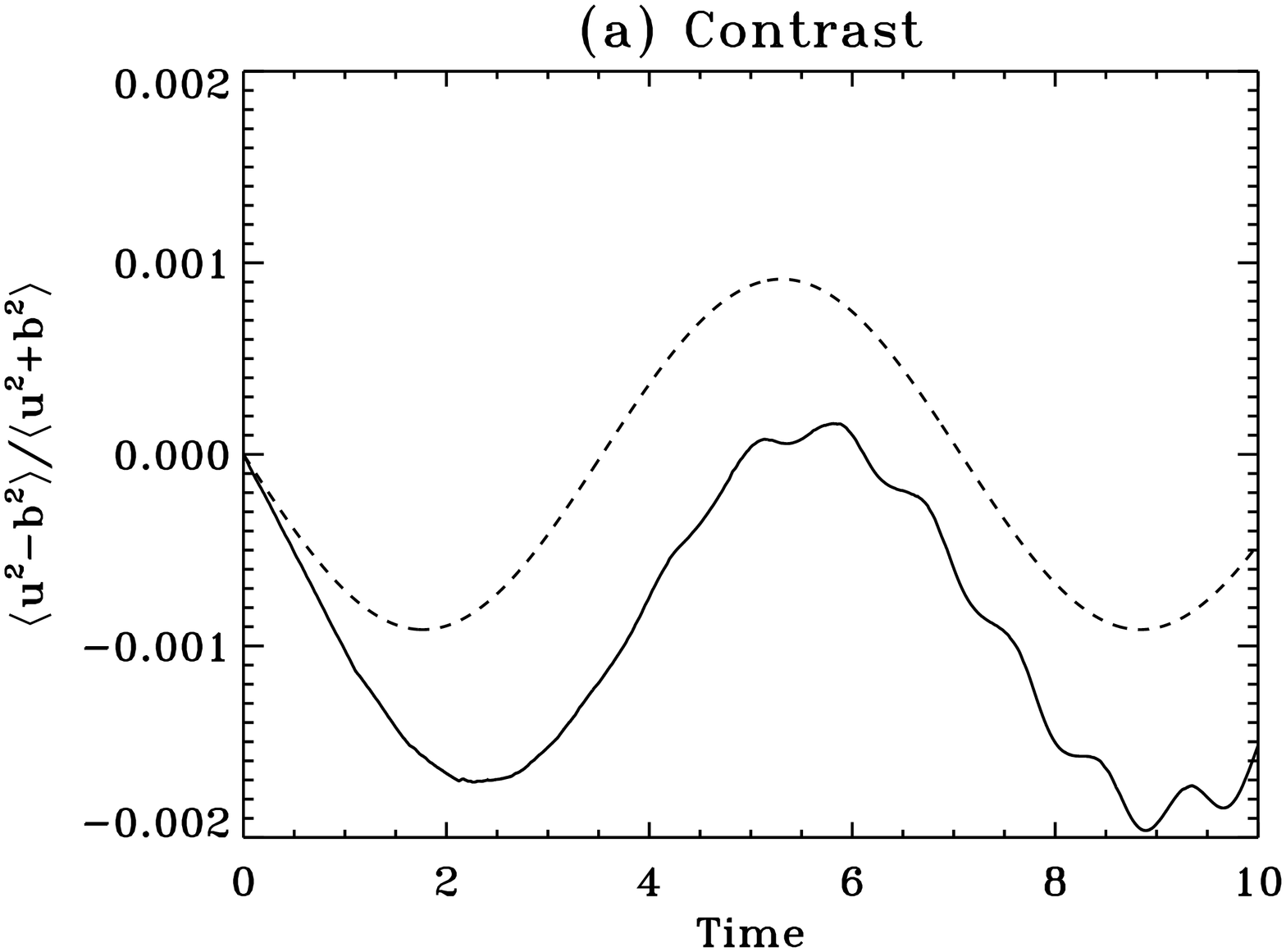,width=6cm,angle=+90}&
\psfig{file=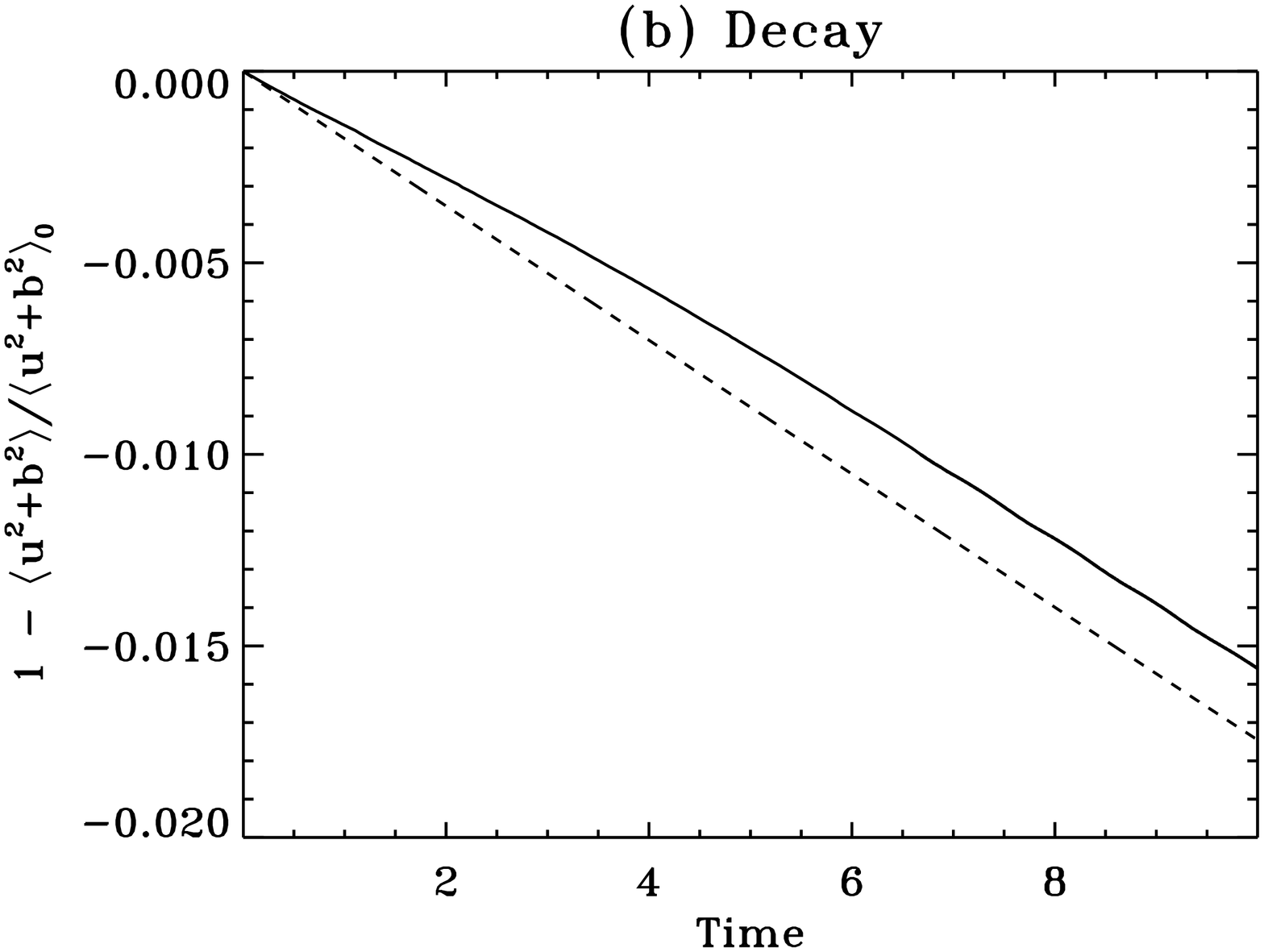,width=6cm,angle=+90}\\
\end{tabular}
} 
\caption{
On the {\it left panel}   (a) we display (solid line) the evolution of
the contrast $<(\Re[\vu])^2-(\Re[\Bb])^2>/<(\Re[\vu])^2+(\Re[\Bb])^2>$
in our standard run as well as (dashed line) the result of equation
(\ref{Cb}) where $\eta-\nu$ is determined from the first time step of
the simulation.   
On the {\it right panel} (b) we display (solid line) the
evolution of the quantity
$1-<(\Re[\vu])^2+(\Re[\Bb])^2>/<2\cos^2(\gb{k.r})>$ in our standard run
as well as (dashed line) the result of equation (\ref{Ca}) where
$\eta+\nu$ is determined from the first time step of the simulation.
 }
\label{check}
\end{figure}
\bibliographystyle{mn}
\bibliography{biblio}

\end{document}